\documentclass[12pt,aps,prd,showpacs,nofootinbib,preprintnumbers,a4paper]{revtex4-1}
\usepackage{amsmath}
\usepackage{amssymb}
\usepackage[pdftex]{graphicx}
\usepackage[english]{babel}
\usepackage{amsfonts}
\usepackage{bbold}
\usepackage{bm}
\usepackage{color}
\usepackage{xcolor}
\usepackage[colorlinks]{hyperref}
\usepackage{cancel}
\usepackage{slashed}
\usepackage{feynmf}

\newcommand{\nn}{\nonumber}

\newcommand{\be}{\begin{eqnarray}}
\newcommand{\ee}{\end{eqnarray}}
\catcode`\@=11
\def\lsim{\mathrel{\mathpalette\@versim<}}
\def\gsim{\mathrel{\mathpalette\@versim>}}
\def\@versim#1#2{\vcenter{\offinterlineskip
\ialign{$\m@th#1\hfil##\hfil$\crcr#2\crcr\sim\crcr } }}
\catcode`\@=12

\newcommand{\Slash}[1]{{\ooalign{\hfil#1\hfil\crcr\raise.167ex\hbo
x{/}}}}

\begin{document}
\allowdisplaybreaks[1]
\widetext

\title{Nambu-Goldstone Dark Matter \\in a  Scale Invariant  Bright   Hidden Sector}
\vspace*{10mm}

\author{Yoshitaka \surname{Ametani}}
\email{ametani@hep.s.kanazawa-u.ac.jp}
\author{Mayumi \surname{Aoki}}
\email{mayumi@hep.s.kanazawa-u.ac.jp}
\author{Hiromitsu \surname{Goto}}
\email{goto@hep.s.kanazawa-u.ac.jp}
\author{Jisuke Kubo}
\email{jik@hep.s.kanazawa-u.ac.jp}
\affiliation{Institute for Theoretical Physics, Kanazawa University, 
Kanazawa 920-1192, Japan}

\preprint{KANAZAWA-15-04}
\vspace*{1cm}

\begin{abstract}
We consider a scale invariant extension of the standard model (SM) with
 a combined breaking of conformal and electroweak symmetry 
in a strongly interacting hidden $SU(n_c)$ gauge sector  with $n_f$ vector-like hidden fermions.
The (pseudo) Nambu-Goldstone   bosons that arise due to
dynamical chiral symmetry breaking are dark matter (DM)  candidates.
 We focus on  $n_f=n_c=3$, where
$SU(3)$ is the largest symmetry group of hidden flavor
which can be explicitly broken into 
either  $U(1) \times U(1)$ or $SU(2)\times U(1)$. 
We study DM properties and discuss
 consistent parameter space for each case.
Because of different  mechanisms of DM annihilation the
consistent parameter space 
in  the case of $SU(2)\times U(1)$  is significantly
different from that of  $SU(3)$ if the hidden fermions have a SM $U(1)_Y$ 
charge of $O(1)$.
 
\end{abstract}

\maketitle

\section{Introduction}
What is the origin of mass?
This is a long-standing question and still remains unsolved  \cite{wilczek}.

The recent discovery of the Higgs particle \cite{Aad:2012tfa,Chatrchyan:2012ufa} may hint how to
go beyond the standard model (SM).
The measured Higgs mass and top quark mass \cite{Agashe:2014kda} are such that
the SM remains perturbative  below the Planck scale 
 \cite{Holthausen:2011aa,Degrassi:2012ry,Bezrukov:2012sa}.
According to Bardeen \cite{Bardeen:1995kv}, ``the SM does not, by itself, have a fine-tuning problem''.
Because the Higgs mass term is the only term,
which breaks scale invariance at the Lagrangian level in the SM, 
we may  ask about
the origin of this mass term.
Mostly scale invariance is hardly broken by quantum anomaly \cite{Callan:1970yg}.
Therefore, a dimensional transmutation can occur at the quantum level,
which can be used to generate a la Coleman-Weinberg \cite{Coleman:1973jx}
the Higgs mass term
 in a classically scale invariant extension of the SM 
 \cite{Fatelo:1994qf}- \cite{Kannike:2015apa}.
 Dynamical chiral symmetry breaking \cite{Nambu:1960xd,Nambu:1961tp} can also be used 
 \cite{Hur:2007uz}-\cite{Heikinheimo:2014xza}.
 The idea is the same as that of technicolor model 
 \cite{Weinberg:1975gm,Susskind:1978ms},
where the only difference is that 
we now allow the existence of fundamental scalars.

 In this paper we consider the latter possibility, in particular
 the model studied in \cite{Hur:2007uz,Hur:2011sv,Heikinheimo:2013fta,Holthausen:2013ota,Kubo:2014ida}.
 In this model  the scale, generated in a QCD-like hidden sector,
is transmitted to the SM sector via a real SM singlet 
scalar   $S$ to trigger spontaneous breaking of electroweak (EW)  gauge  symmetry
\cite{Hur:2007uz,Hur:2011sv} (see also \cite{Kubo:2014ova}). Moreover, 
due to the dynamical  chiral symmetry breaking   in the hidden sector
there exist  Nambu-Goldstone (NG)  bosons, which are massive,
because the coupling $y$ of 
 $S$ with the hidden sector fermions breaks explicitly  chiral symmetry.
Therefore,  the mass scale of the NG bosons,
which are dark matter (DM) candidates,
is not independent (as it is not the case in the most of DM models);
it is smaller than the hidden sector scale,
which is in the TeV region unless the coupling
$y$ is very small, i.e. $\lsim O(10^{-4})$.

As in \cite{Holthausen:2013ota,Kubo:2014ida} we
employ the Nambu-Jona-Lasinio  (NJL) theory \cite{Nambu:1960xd,Nambu:1961tp}
as an low-energy effective theory
of the hidden sector and base our calculations
on the self-consistent mean field (SCMF) approximation 
\cite{Hatsuda:1994pi,Kunihiro:1983ej} of the  NJL theory, 
which  is briefly outlined in Sect.~III.
In \cite{Holthausen:2013ota,Kubo:2014ida}  the maximal global flavor symmetry
$SU(3)_V$ (along with a $U(1)_V$) has been assumed.
In this paper we  relax this assumption  and consider  in detail the cases, in which 
$SU(3)_V$ is broken into its subgroups.
We  find in Sect.~IV  that the consistent parameter space can be 
considerably extended if $SU(3)_V$ is broken 
to its subgroup $SU(2)_V\times U(1)_{\tilde{B}'}$.
The main reason is that, if  $SU(3)_V$ is broken,
a new mechanism for the DM annihilation, inverse DM conversion,
becomes operative at finite temperature:
A pair of lighter DM particles annihilate into 
 a pair of heavier (would-be) DM particles,
 which subsequently decay into SM particles
 (mainly into two $\gamma$s).

 Before we discuss the DM phenomenology of the model, we 
 develop an effective  theory for DM interactions
 (a linear sigma model) in the framework of 
 the SCMF approximation of the NJL theory.
Using the effective theory we compute the DM relic abundance
and analyze the direct and indirect DM detection possibilities in Sect.~IV.
 Sect.~V is devoted to Conclusion, and in Appendix A 
 we give explicitly the NJL Lagrangian in the SCMF approximation
 in the case that
 $SU(3)_V$ is broken into $U(1)_{\tilde{B}'}\times U(1)_{\tilde{B}}$.
In Appendix B the inverse DM (mesons for QCD) propagators 
and also how the NJL parameters are fixed  can be found.
The one-loop integrals that are used in our
calculations are collected in Appendix C.

\section{The model}
We consider  a classically scale invariant extension of the SM studied in 
\cite{Hur:2007uz,Hur:2011sv,Heikinheimo:2013fta,Holthausen:2013ota,Kubo:2014ida} \footnote{See 
also \cite{Strassler:2006im}.}
which consists of a hidden $SU(n_c)_H$ gauge sector 
coupled via a real singlet scalar $S$ to  the SM.
The hidden sector Lagrangian of the model 
is written as 
\be
{\cal L}_{\rm H}
&=-\frac{1}{2}\mbox{Tr}~F^2+
\mbox{Tr}~\bar{\psi}(i\gamma^\mu \partial_\mu +
g_H \gamma^\mu G_\mu +g'  Q \gamma^\mu B_\mu-
y S)\psi~,
\label{eq:LH}
\ee
where $G_\mu$ is the gauge field for the hidden QCD,
$B_\mu$ is the $U(1)_Y $ gauge field, i.e.
\be
B_\mu &=&\cos \theta_W A_\mu-\sin\theta_W Z_\mu~,~
g'=e/\cos \theta_W ~,
\ee
and
the $n_f$ (Dirac) fermions $\psi_i~(i=1,\dots,n_f)$ in the hidden sector 
belong to the fundamental representation
of  $SU(n_c)_H$.
The trace in (\ref{eq:LH}) is taken over the flavor as well as the color indices.
The hidden  fermions carry a common $U(1)_Y$ charge $Q$, implying that
they contribute only to
$\Pi_{YY}$ of the gauge boson self-energy diagrams so that
 the $S, T, U$ parameters remain unchanged.
The ${\cal L}_{\mathrm{SM}+S}$ part of the total Lagrangian ${\cal L}_T ={\cal L}_{\rm H}+{\cal L}_{\mathrm{SM}+S} $ contains the 
SM gauge and Yukawa  interactions
along with the scalar potential
\be
V_{\mathrm{SM}+S}
&=&
\lambda_H ( H^\dag H)^2+
\frac{1}{4}\lambda_S S^4
-\frac{1}{2}\lambda_{HS}S^2(H^\dag H)~,
\label{eq:VSM}
\ee
where  $H^T=( H^+ ~,~(h+iG)\sqrt{2}  )$ is the SM Higgs doublet field,
with $H^+$ and $G$ as the would-be Nambu-Goldstone fields
 \footnote{This classically scale invariant 
 model is perturbatively renormalizable, and 
 the Green's functions are infrared finite
 \cite{Lowenstein:1975rf,Poggio:1976qr}.}.
The basic mechanism to trigger the  EW  symmetry breaking 
is very simple:
The  non-perturbative effect of dynamical 
chiral symmetry breaking in the hidden sector
 generates a robust scale which is transferred into the SM sector through the real singlet $S$.
Then the mass term for the Higgs potential is 
generated via the Higgs portal term in (\ref{eq:VSM}), where
the ``$-$'' in front of the positive $\lambda_{HS}$ is an assumption.

 \subsection{Global Symmetries}
 The Yukawa coupling of the hidden fermions with the singlet $S$ breaks 
explicitly chiral symmetry. Therefore, in the  limit of the vanishing Yukawa coupling matrix 
 $y_{ij}$ the global symmetry $SU(n_f)_L\times SU(n_f)_R
 \times U(1)_V\times U(1)_A$
 is present at the classical level, where $U(1)_A$ is broken by anomaly at the quantum level down to its discrete subgroup $Z_{2n_f}$, and 
the unbroken $ U(1)_V$ ensures the conservation of the hidden baryon number.
 The non-abelian part of the chiral symmetry $SU(n_f)_L\times SU(n_f)_R$
 is broken dynamically down to its diagonal subgroup $SU(n_f)_V$ by the non-vanishing
 chiral condensates $ \langle\bar{\psi}_i \psi_i\rangle$, implying the existence of
 $n_f^2-1$ NG bosons $\phi_a~(a=1,\dots,n_f^2-1)$.
 In the $n_f=3$ case the NG bosons are like the mesons in the real hadron world:
 \be
\tilde{ \pi}^0 &=& \phi_3~,~
 \tilde{\pi}^\pm = (\phi_1\mp i\phi_2)/\sqrt{2}~,~\nn\\
 \tilde{K}^\pm & = & (\phi_4\mp i  \phi_5)/\sqrt{2}~,
 ~\tilde{K}^0(\bar{\tilde{K}}^0) =
  (\phi_6+(-) i \phi_7)/\sqrt{2}~,~ \tilde{\eta}^8 = \phi_8~,
  \label{mesons}
  \ee
where $\tilde{\eta}^8$ will mix with $\tilde{\eta}^0$ to form the mass eigenstates
$\tilde{\eta}$ and $\tilde{\eta}'$.
(The $\tilde{}$ should avoid the confusion with the real mesons $\pi^0$ etc.)
 
In the presence of the Yukawa coupling the chiral symmetry is explicitly broken;
this is the only coupling which breaks the chiral symmetry  explicitly.
Because of this coupling the NG bosons become massive.
An appropriate chiral rotation of $\psi_i$ can diagonalize
 the Yukawa coupling matrix:
 \be
 y_{ij}=y_i\delta_{ij}~(y_i \geq 0)
 \label{yukawa}
 \ee
 can be assumed without loss of generality,
which implies that $U(1)^{n_f-1}$ corresponding to the elements of
the Cartan subalgebra of $SU(n_f)$  are unbroken.
We assume that none of $y_i$ vanishes so that all the NG bosons are massive.
If two $y_i$s are the same, say $y_1=y_2$, one $U(1)$ is promoted to
an $SU(2)$. Similarly,
if three $y_i$s are the same,   a product group $U(1)\times U(1)$ is promoted to
an $SU(3)$, and so on.
  In addition to these symmetry groups, there exists a discrete $Z_4$,
 \be
 Z_4 &:& \psi_i \to  \left(\exp i (\pi/2)\gamma_5\right)  \psi_i
 =i\gamma_5 \psi_i~\mbox{and}~S \to -S~.
 \label{z4}
 \ee
This discrete symmetry is anomalous for odd $n_f$, because 
the chiral transformation in (\ref{z4}) is an element
of the anomalous $U(1)_A$. If $n_f$ is even,  then the chiral transformation is an
element of the anomaly-free subgroup  $Z_{2n_f}$ of $U(1)_A$.
Needless to say that  this $Z_4$ is broken by 
a non-vanishing vacuum expectation value (VEV) of $S$,
which is essential to trigger the  EW   gauge symmetry breaking.

 \subsection{ Dark Matter Candidates}
The NG bosons, which arise due to the 
dynamical chiral symmetry breaking in the hidden sector,
are good DM candidates, because they are neutral and their interactions with the SM
part start to exist at the one-loop level so that they are weak.   
However, not all NG bosons can be  DM, because their stability  depends on 
the global symmetries that are in tact.
In the following we consider the  case for $n_f=3$, which can be simply
extended to  an arbitrary $n_f$.
For $n_f=3$
there are three possibilities of the global symmetries:
\be
 & & \mbox{(i)}~U(1)_{\tilde B'}\times U(1)_{\tilde B}~\mbox{if}~
 y_1\neq y_2\neq y_3~, \label{u1u1}\\
& &\mbox{(ii)} ~ SU(2)_V\times U(1)_{\tilde B}~
\mbox{if}~
 y_1=y_2\neq y_3~, \label{su2u1}\\
& & \mbox{(iii)}  ~ SU(3)_V~\mbox{if}~
 y_1=y_2= y_3~ ,\label{su3}
\ee
where we have suppressed
$U(1)_V$ which always exists, and the case (iii) has been treated in detail
in \cite{Kubo:2014ida}. Without loss of generality we can assume that the 
elements of the Cartan subalgebra corresponding to
$U(1)_{\tilde B'}$ and $U(1)_{\tilde B}$ are
\be
{\tilde B'}&=&\left(\begin{array}{ccc}
1 & 0 & 0\\
0 & -1& 0\\
0 & 0 & 0\\
\end{array}\right)~,~{\tilde B} = \left(\begin{array}{ccc}
1 & 0 & 0\\
0 & 1& 0\\
0 & 0 & -2\\
\end{array}\right)~.\label{BBp}
\ee
In Table I we show the NG bosons for $n_f=3$ with their quantum numbers.
As we can see from Table I the NG bosons $\tilde{\pi}^0$ and $\tilde{\eta}^8$ are
unstable for the case (i) and in fact can decay into two $\gamma$s, while
for the  case (ii) only $\tilde{\eta}^8$ is unstable.
Whether the stable NG bosons can be realistic DM particles
is a dynamical question, which we will address later on.

\begin{table}
\caption{\footnotesize{The NG bosons and DM candidates for $n_f=3$.}}
\begin{center}
\begin{tabular}{|c|ccccccccccccc|} 
\hline
 & $\tilde{\pi}^0$ & $\tilde{\pi}^+$ & $\tilde{\pi}^-$  &  & $\tilde{K}^0$ 
&  & $\tilde{K}^+$  & & $\tilde{K}^-$ &  &  $\tilde{\bar{K}}^0$ 
&  &$\tilde{\eta}^8$ \\
\hline
$U(1)_Y$ charge& $0$ & $0$ & $0$ & & $0$ &
 & $0$ &  & $0$ &  &  $0$ 
&  &$0$ \\
\hline
${\tilde B'}$ & $0$ & $2$ & $-2$ & & $-1$ &
 & $1$ &  & $-1$ &  &  $1$ 
&  &$0$ \\
\hline
${\tilde B}$ & $0$ & $0$ & $0$  & & $3$ &
 & $3$ &  & $-3$ &  &  $-3$ 
&  &$0$ \\
\hline
$SU(2)_V$ &  & ${\bf 3}$ &   & &  &
${\bf 2}$ &  & &  &${\bf 2}$ &   
&  &${\bf 1}$ \\
\hline
$SU(3)_V$ &  &   &   & &  &
 & ${\bf 8}$ & &  &  &   
&  & \\
\hline
\end{tabular}
\end{center}
\label{contents}
\end{table}

 \subsection{Perturbativity and stability of the scalar potential at high energy}
 
 Before we discuss the non-perturbative effects, we consider briefly the perturbative part
 at high energies, i.e. above the scale of the dynamical chiral symmetry breaking
 in the hidden sector. As explained in the Introduction, it is essential for our scenario
 of explaining the origin of the EW scale to work that 
  the scaler potential is unbounded below and the theory remains perturbative 
 (no Landau pole) below the Planck scale.
 So, we require:  
\be
& & ~4\pi > \lambda_H,\lambda_S >0,~
4\pi > \lambda_{HS}> 0~,|y|^2 <4\pi,
\label{cond1}\\
&  &~2\sqrt{\lambda_H \lambda_S}-\lambda_{HS}>0~.\label{cond2}
\ee
In the following discussion we assume that the perturbative regime 
(of the hidden sector) starts  around $q_0=1$ TeV and $g_H^2 (q_0)/4\pi
\simeq1$.
 Although in this model the Higgs mass depends mainly on two parameters, $\lambda_H$ and $\lambda_{HS}$, lowering $\lambda_H(q_0)<0.13$ will destabilize the Higgs potential while increasing $\lambda_H(q_0)> 0.14$ will require a larger mixing with $S$, which is strongly constrained. Therefore, we consider the RG running of the couplings with
$\lambda_H(q_0)$ fixed at $ 0.135$ and rely on one-loop approximations.
In the case that the hypercharge $Q$ of the hidden fermions is different
 from zero,
these fermions contribute to the renormalization group (RG) running of the $U(1)_Y$ gauge coupling considerably.
We found that $Q \lsim 0.8$ should be satisfied for $g'$ to remain perturbative
below the Planck scale. 

\begin{figure}
\includegraphics[width=0.6\textwidth]{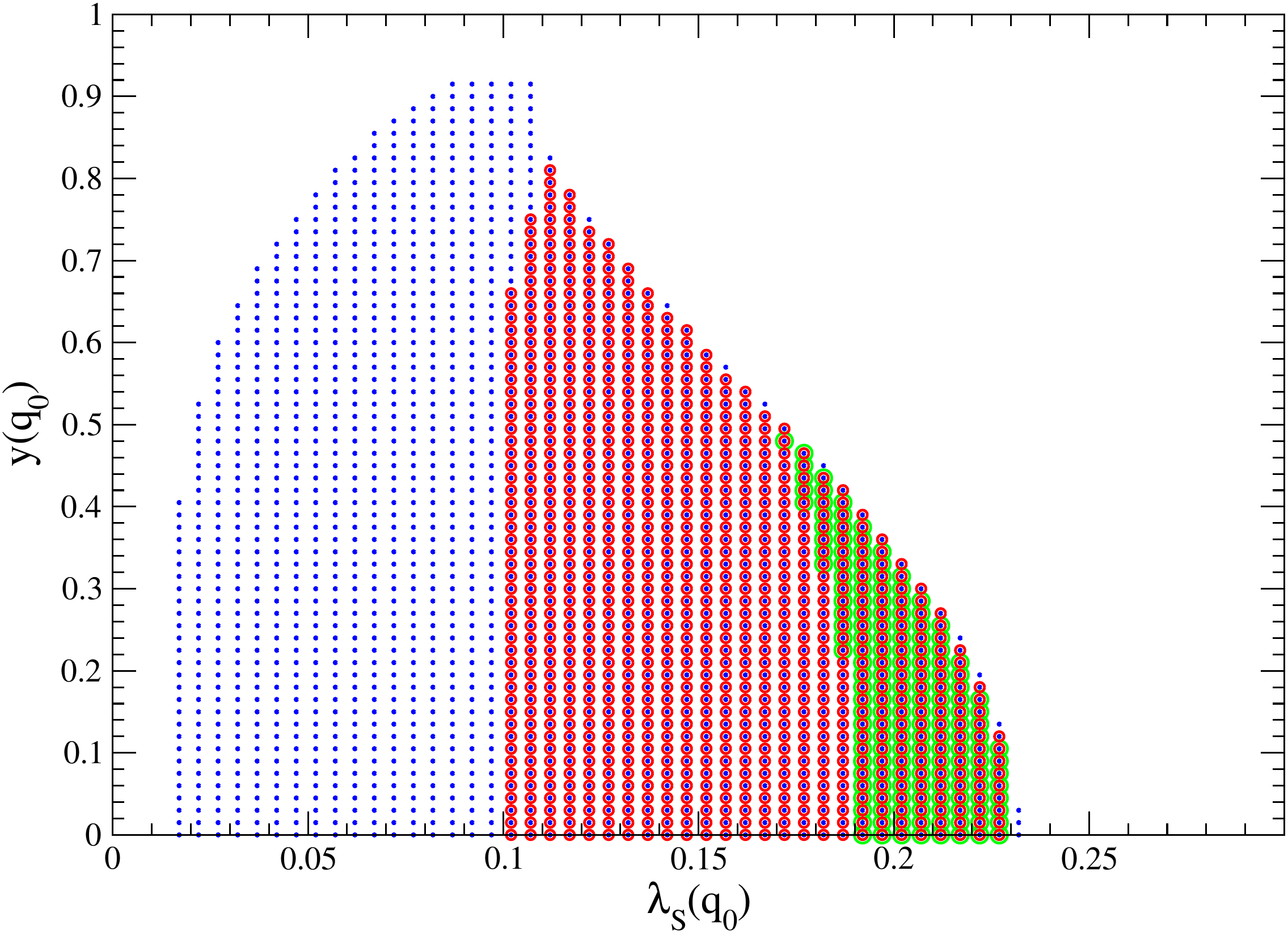}
\caption{ \label{Ls-y}
\footnotesize
Stability constraint.
The allowed area in the $\lambda_S-y$ plane for 
different values of $\lambda_{HS}(q_0)$
with $\lambda_H(q_0)$ fixed at $ 0.135$ ($q_0=1$ TeV)
is shown, where  we have used $Q=1/3$
and assumed the $SU(3)_V$ flavor symmetry defined 
in (\ref{su3}).
The green circles,  red circles and blue points stand for
$\lambda_{HS}(q_0)=0.1, 0.06$ and $0.02$.
}
\end{figure}

Because of (\ref{cond2}) the range of $\lambda_{S}$ is constrained 
for a given $\lambda_{HS}$ and $\lambda_{H}$: The larger $\lambda_{HS}$ is, the larger   $\lambda_{S}$ has to be. But there is an upper limit for $\lambda_{S}(q_0)$ 
because of perturbativity .
In Fig.~\ref{Ls-y} we show the allowed area in the $\lambda_S-y$ plane for 
different values of $\lambda_{HS}(q_0)$
with $\lambda_H(q_0)$ fixed at $ 0.135$
in the $SU(3)_V$ case (\ref{su3}), i.e.
$y=y_1=y_2=y_3$
\footnote{The same analysis has been performed in \cite{Kubo:2014ida}, but without
including the constraint (\ref{cond2}).}.
The green circles, red circles and blue points stand for
$\lambda_{HS}(q_0)=0.1, 0.06$ and $0.02$.
There will be no allowed region for $\lambda_{HS}(q_0) \gsim 0.12$.
We have used $Q=1/3$, but
the allowed area does not depend very much on $Q$ as long as $Q < 0.8$
is satisfied (which ensures perturbativity of the $U(1)_Y$ gauge coupling).
 If $SU(3)_V$ is broken, then 
the vertical axis in Fig.~\ref{Ls-y} represents the largest among
$y_i$s.

\section{Nambu-Jona-Lasinio Method}
\subsection{NJL Lagrangian in a mean-field approximation }

Following \cite{Holthausen:2013ota} 
we replace the high energy Lagrangian ${\cal L}_{H}$ in (\ref{eq:LH}) by the NJL Lagrangian
\be
{\cal L}_{\rm NJL}&=&\mbox{Tr}~\bar{\psi}(i\gamma^\mu\partial_\mu 
+g' Q \gamma^\mu B_\mu-y S)\psi+2G~\mbox{Tr} ~\Phi^\dag \Phi
+G_D~(\det \Phi+h.c.)~,
\label{eq:NJL10}
\ee
where
\be
\Phi_{ij}&=& \bar{\psi}_i(1-\gamma_5)\psi_j=
\frac{1}{2}\sum_{a=0}^{n_f^2-1}
\lambda_{ji}^a \mbox{Tr}~\bar{\psi}\lambda^a(1-\gamma_5)\psi~,
\ee
and $\lambda^a (a=1,\dots, n_f^2-1)$ are the Gell-Mann matrices with
$\lambda^0=\sqrt{2/3}~{\bf 1}$.
The effective Lagrangian ${\cal L}_{\rm NJL}$ has
three dimensional parameters
$G,G_D$ and the cutoff $\Lambda$, which have canonical dimensions of 
$-2$, $-5$ and $1$,  respectively.
Since the original Lagrangian ${\cal L}_{H}$ has only one independent scale,
the parameters $G,G_D$ and  $\Lambda$
are not independent.
We restrict ourselves to $n_c=n_f=3$, because
in this case these parameters, up-to an overall scale,
can be approximately fixed from  hadron physics 
\cite{Kunihiro:1983ej,Hatsuda:1994pi}.
The six-fermi interaction in (\ref{eq:NJL10}) is present due to chiral anomaly
of the axial $U(1)_A$ and is invariant under $Z_6$,
so that  the NJL Lagrangian (\ref{eq:NJL10}) has the same global symmetry
as the high energy Lagrangian  (\ref{eq:LH}).
Furthermore, as we  mentioned in Sect.~II A, we can assume without loss of generality that  the Yukawa coupling matrix $y$  is diagonal (see (\ref{yukawa})).
To deal with the non-renormalizable Lagrangian (\ref{eq:NJL10}) 
 we employ \cite{Holthausen:2013ota}
the SCMF approximation which has been 
 intensely studied by Hatsuda and Kunihiro
\cite{Kunihiro:1983ej,Hatsuda:1994pi} for hadron physics. 
The NJL parameters for the hidden QCD is then obtained by the upscaling of the actual values of $G,G_D$ and the cutoff $\Lambda$ from QCD hadron physics.
 That is, we assume that the dimensionless combinations
 \be
G^{1/2} \Lambda &=&1.82~,~(-G_D)^{1/5} \Lambda =2.29~,
\label{G-GD}
\ee
which are  satisfied for hadrons, remain unchanged for
a higher scale of $\Lambda$.

Below we briefly outline  the SCMF approximation.
We go  via a Bogoliubov-Valatin transformation from the perturbative  vacuum  to the ``BCS'' vacuum,
which we simply denote by
$|0\rangle$.
This vacuum is so defined that the mesons (mean fields)
are collected in  the VEV of the chiral bilinear:
\be
 \varphi &  \equiv & \langle0 \vert\bar{\psi}(1-\gamma_5)\lambda^a \psi
 \vert 0 \rangle=
-\frac{1}{4G}  \left(\mbox{diag}(\tilde{\sigma}_1,
\tilde{\sigma}_2,\tilde{\sigma}_3)+i(\lambda^a)^T \phi_a\right)~,
\label{eq:varphi}
 \ee
where we denote the pseudo NG boson fields  after spontaneous chiral symmetry breaking by $\phi_a$. 
The dynamics of the hidden sector creates a nonvanishing chiral condensate
 $\langle 0 \vert \bar{\psi}_i \psi_i \vert 0 \rangle$
 which is nothing but  $-\langle\tilde{\sigma}_i \rangle/4 G$.
 The actual value of $\langle\tilde{\sigma}_i \rangle$ can be obtained
through the minimization of the scalar potential, as we describe shortly.
In the SCMF approximation one splits up the NJL Lagrangian (\ref{eq:NJL10}) into the sum
\be
\mathcal{L}_{\rm NJL} &= &\mathcal{L}_{0}+\mathcal{L}_{I}~,
\ee
where  $\mathcal{L}_{I}$ is 
normal ordered 
(i.e. $\langle 0\vert \mathcal{L}_{I}\vert 0\rangle =0$), and $\mathcal{L}_{0}$ contains at most fermion bilinears
which are not normal ordered.
At the non-trivial lowest order only ${\cal L}_0$ is relevant
for the calculation of  the effective potential, the
DM mass and the DM interactions.
The explicit form for $\mathcal{L}_{0}$ can be found in Appendix A.
The effective potential can be obtained by integrating out the hidden
fermion fields in the BCS vacuum. At the one-loop level we find
\be
V_{\rm NJL}(\tilde{\sigma}_i,S)
&  &= \frac{1}{8G}\sum_{i=1,2,3}\tilde{\sigma}_i^2-
\frac{G_D}{16G^3}\tilde{\sigma}_1\tilde{\sigma}_2\tilde{\sigma}_3
-\sum_{i=1,2,3} n_c I_V(M_i)~,
\label{eq:Vnjl}
\ee
where $I_V(m)$ is given in Eq.~(\ref{vac}), and
the constituent fermion masses $M_i$ are given by
\be
M_{i}& =&\tilde{\sigma}_{i}+y_{i} S-\frac{G_D}{8G^2}
\tilde{\sigma}_{i+1}\tilde{\sigma}_{j+2},~
\label{eq:cmass0}
\ee
where $\tilde{\sigma}_{4}=\tilde{\sigma}_{1}$ and
$\tilde{\sigma}_{5}=\tilde{\sigma}_{2}$.
Once the free parameters of the model 
$y_i,\lambda_H,\lambda_{HS},\lambda_S$ 
are given,
the VEVs of $\tilde{\sigma}_i$ and $S$ can be determined through
the minimization of the scalar  potential
$V_{\rm SM + S}+V_{\rm NJL} $, 
where  $V_{\rm SM + S}$ is defined in (\ref{eq:VSM}).
After the  minimum of the scalar potential is fixed, 
the mass spectrum for the CP-even particles $h,S$ and 
$\tilde{\sigma}$ as well as  the DM candidates with their
properties are obtained.

\subsection{The value of $y$ and hidden Chiral Phase Transition}
The Yukawa coupling in (\ref{eq:LH}) violates explicitly chiral symmetry and  plays
a similar role as the current quark mass in QCD. 
It is well known that the nature of 
 chiral phase transition in QCD depends  on the value 
  of the current quark mass.
Therefore, it is expected that the value of $y$ strongly influences 
the nature of the chiral phase transition
in the hidden sector, which has been confirmed in 
\cite{Holthausen:2013ota}. The hidden chiral phase transition occurs  above the EW  phase transition, where
 the nature of the EW phase transition is not known yet.
In the following discussions, we restrict ourselves to
\be
y\lsim 0.006~,
\label{y-range}
 \ee
because in this case 
the hidden chiral phase transition is a strong first order transition
\cite{Holthausen:2013ota} and
can produce gravitational wave back ground 
\cite{Witten:1984rs,Hogan:1986qda}, which could be observed 
by future experiments such as
 Evolved Laser Interferometer Space Antenna (eLISA)
experiment \cite{AmaroSeoane:2012km}. Needless to say that the smaller is $y$, the better
is the NJL approximation to chiral symmetry breaking.

\section{Dark Matter Phenomenology}

\subsection{Dark Matter Masses}
Our DM candidates are the pseudo NG bosons,
which occur due to the dynamical
 chiral symmetry breaking in the hidden sector.
They are CP-odd scalars,
and  their masses are generated at one-loop in the SCMF approximation
as the real meson masses, where we here, too, restrict ourselves to
$n_c=n_f=3$.
Therefore, their inverse propagators can be calculated 
in a similar way as in  the QCD case, which 
is given in Appendix B.

\begin{figure}
\includegraphics[width=0.4\textwidth]{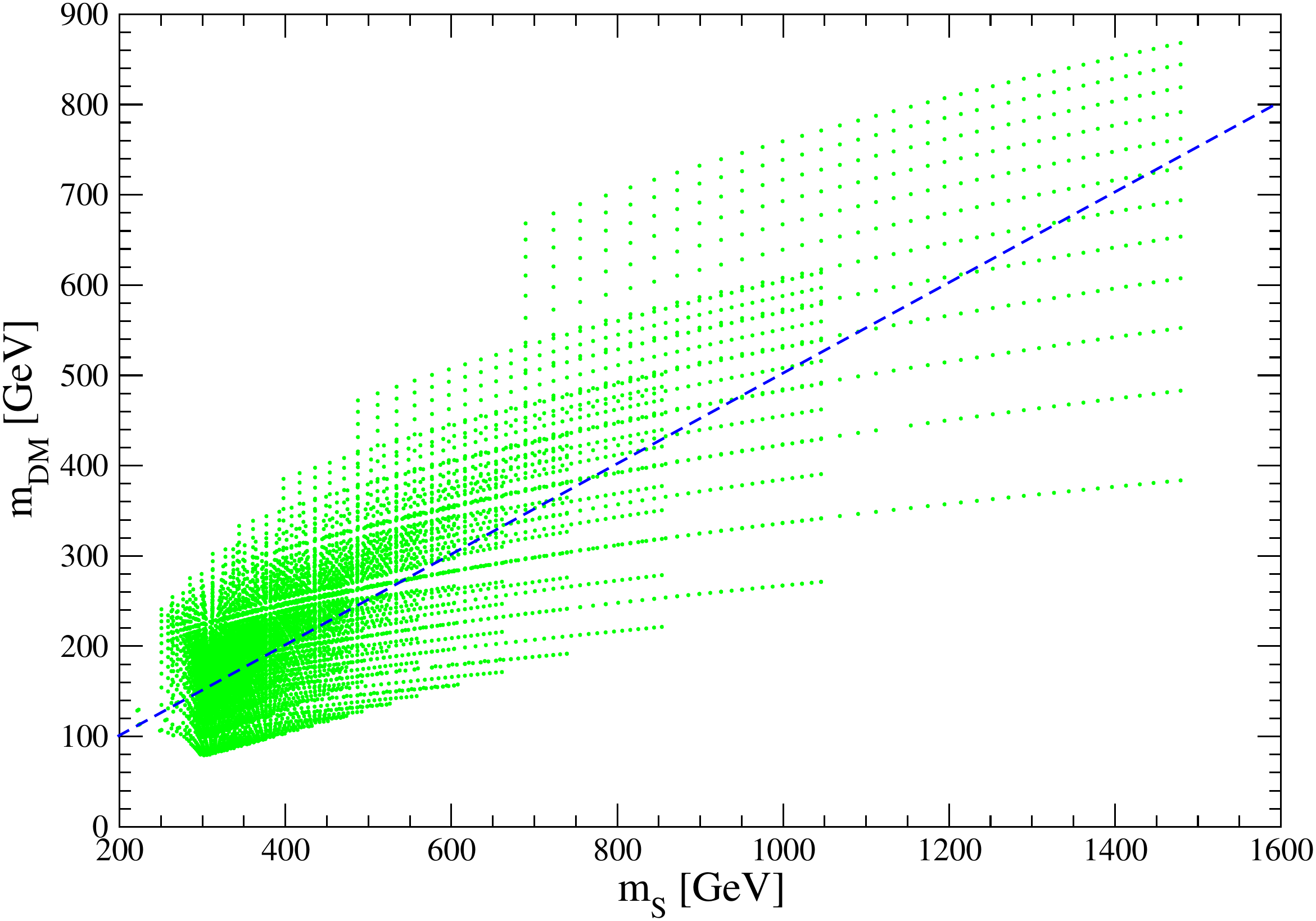}\includegraphics[width=0.4\textwidth]{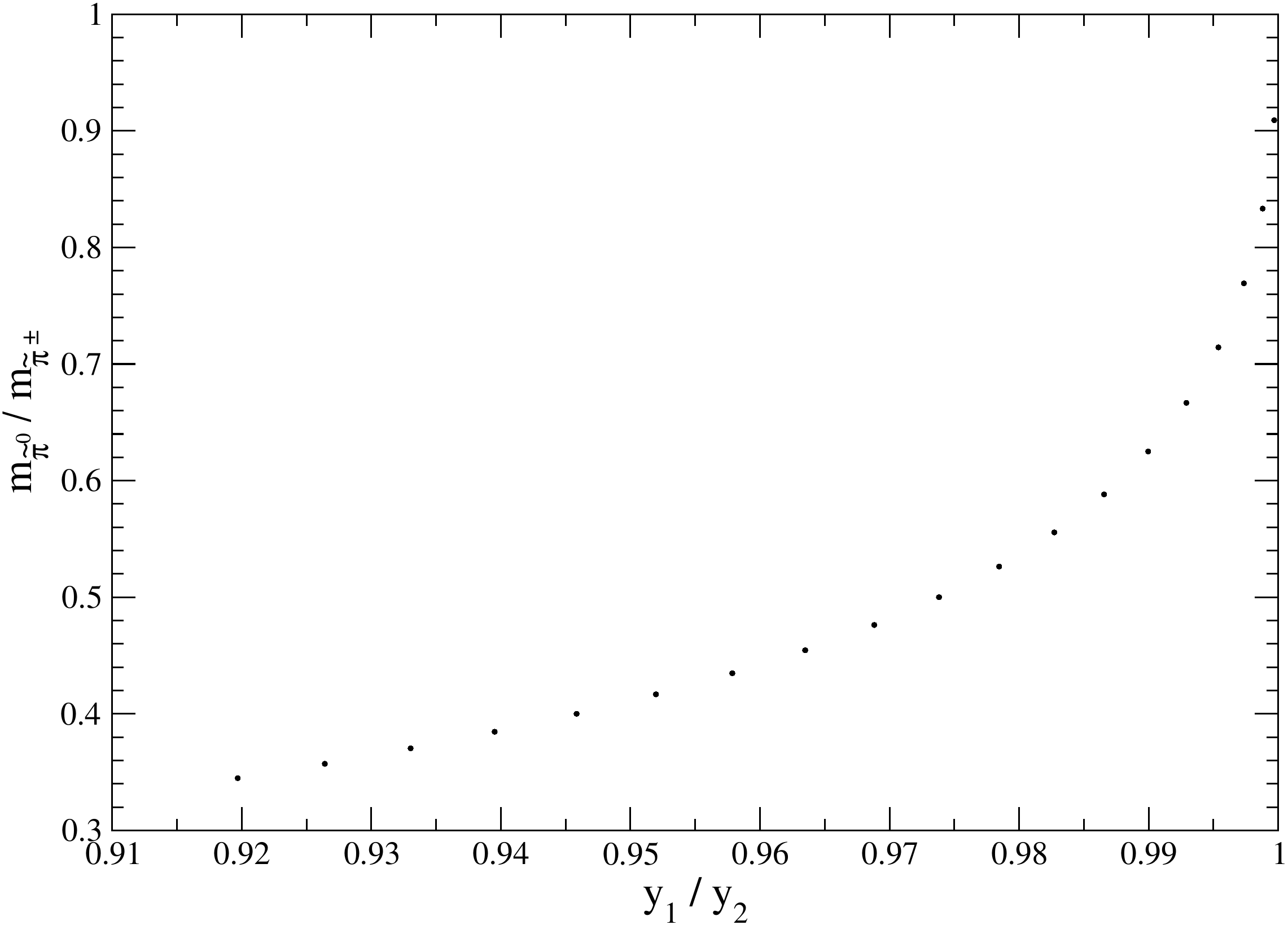}
\caption{ \label{mpi-ratio}
\footnotesize
Left: The area in the $m_S$-$m_{\rm DM}$ plane
for $0.0005< y_1=y_2=y_3 <0.006$ in the $SU(3)_V$ case
(\ref{su3}), where
$m_{\rm DM}=m_{\tilde{\pi}^0}=m_{\tilde{\pi}^\pm}
=m_{\tilde{K}^0}
=m_{\tilde{K}^\pm}
=m_{\tilde{\eta}}$.
The ratio $m_S/m_{\rm DM}=2$ is satisfied on the blue dashed line,
on which the resonance condition in  the s-channel diagram 
for the DM annihilation (Fig.~\ref{pp-to-SMSM}) is satisfied.
Right: The ratio $m_{\tilde{\pi}^0}/m_{\tilde{\pi}^\pm}$
versus $y_1/y_2$ in the $U(1)_{\tilde{B}'}\times U(1)_{\tilde{B}}$ 
case (\ref{u1u1}), 
where $y_1$ and $y_3$ are fixed at
$0.002$ and $0.006$, respectively.
The constraints  imposed on $\lambda_H,
\lambda_{HS}$ and $\lambda_S$
are such that 
we obtain a correct Higgs mass and
the perturbativity (\ref{cond1}) as well as stability (\ref{cond2})  constraints 
are satisfied.
}
\end{figure}

First we consider the $SU(3)_V$ case (\ref{su3}) to obtain the DM mass
$m_{\rm DM}$
\footnote{Since $SU(3)_V$ is unbroken, all the DM particles have the mass
which is denoted by $m_{\rm DM}$ here.} and the mass of the singlet $m_S$
for $0.001 \lsim y_1= y_2= y_3 \lsim 0.006$.
In Fig.~\ref{mpi-ratio} (left) we show the area in the
 $m_S$-$m_{\rm DM}$ plane,
in which  we obtain a correct  Higgs mass,
while imposing the perturbativity (\ref{cond1})  as well as  stability 
(\ref{cond2}) constraints. 
The upper  limit of $m_{\rm DM}$ for a given $m_S$ 
is due to the upper limit of 
the Yukawa coupling (see (\ref{y-range})),
while its lower limit comes from the 
lower limit of  the Yukawa coupling, which is taken to be $ 0.0005$ here.
The upper limit for $m_S$ is dictated by the upper limit of $\lambda_S$,
which is fixed by the perturbativity and stability constraints
 (\ref{cond1}) and  (\ref{cond2}).
 The lowest value of  $m_S$, $250$ GeV,
 comes from the lowest value of $\lambda_S$,
 which is set  at $ 0.05$ here.
 If $SU(3)_V$ is only slightly broken, the DM mass will not change very much.

We next consider the $U(1)_{\tilde{B}'}\times U(1)_{\tilde{B}}$ case (\ref{u1u1}).
We may assume without loss of generality that the hierarchy $y_1 < y_2 < y_3$
is satisfied.
In  Fig.~\ref{mpi-ratio} (right) we show the ratio $m_{\tilde{\pi}^0}/m_{\tilde{\pi}^\pm}$
versus $y_1/y_2$, where we have fixed $y_1$ and $y_3$ at
$0.002$ and $0.006$, respectively.
We can conclude from Fig.~\ref{mpi-ratio} (right) that $\tilde{\pi}^0$ is the lightest among the
pseudo NG bosons and the ratio $m_{\tilde{\pi}^0}/m_{\tilde{\pi}^\pm}$
does not practically depend on the scalar couplings $\lambda_H,
\lambda_{HS}$ and $\lambda_S$.
The $SU(2)_V$ case (\ref{su2u1}) can be realized if two 
of $y_i$ are the same. There are two independent possibilities:
 (a)~ $y_1=y_2 < y_3$,
and (b)~ $y_1 < y_2 =y_3$.
The mass spectrum for the case (b) is similar to that for 
 the $U(1)_{\tilde{B}'}\times U(1)_{\tilde{B}}$ case.
 In particular, $\tilde{\pi}^0$ is the lightest among the
pseudo NG bosons.
As for the case (a) the mass hierarchy 
\be
m_{\tilde{\pi}} &=& m_{\tilde{\pi}^0} 
 =m_{\tilde{\pi}^\pm} < m_{\tilde{K}} =m_{\tilde{K}^0}=m_{\tilde{K}^\pm}
< m_{\tilde{\eta}}
\label{hrcy}
\ee
is always satisfied.

The different type of the DM  mass spectrum  will have an important consequence
when discussing the DM relic abundance.

\subsection{Effective interactions for DM Decay and Annihilations}

As discussed in Sect. II B, if the $SU(3)_V$ flavor symmetry is broken 
to $U(1)_{\tilde{B'}}\times U(1)_{\tilde{B}}$, 
there will be two real decaying would-be DM particles
$\tilde{\eta},\tilde{\pi}^0$, and  three pairs  of complex DM particles
$(\tilde{K}^0,\tilde{\bar{K}}^0), \tilde{K}^\pm$ and $ \tilde{\pi}^\pm$.
Here we will  derive effective interactions for these DM fields by integrating out
the hidden fermions at the one-loop order.
The one-loop integrals and their lowest order expressions
of   expansion in the external momenta
in the large $\Lambda$ limit  are given in Appendix B.
Except for the $\phi\mbox{-}\phi\mbox{-}\gamma$ 
and $\phi\mbox{-}\gamma\mbox{-}\gamma$ interactions,
we assume  $SU(2)_V$ flavor symmetry, i.e.
\be
\langle\sigma_1 \rangle &=& \langle\sigma_2 \rangle,~
M_1=M_2~,Z_{\tilde{K}}=Z_{\tilde{K}^\pm}
=Z_{\tilde{K}^0}~,~
Z_{\tilde{\pi}}=Z_{\tilde{\pi}^\pm}=
Z_{\tilde{\pi}^0}~,\label{su2}
\ee
where $Z$s are the wave function renormalization constants given in (\ref{wave}).
This is because, we have to assume at least $SU(2)_V$
for a realistic parameter space as we will see.

\begin{figure}
\includegraphics[width=0.3\textwidth]{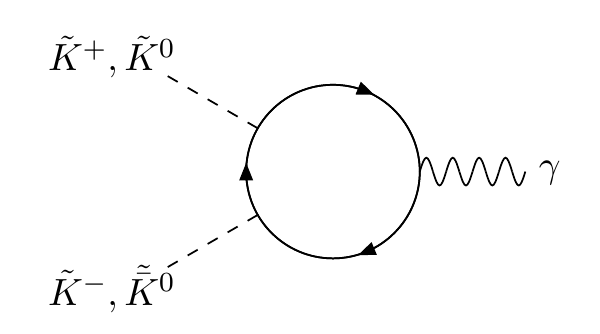}
\includegraphics[width=0.29\textwidth]{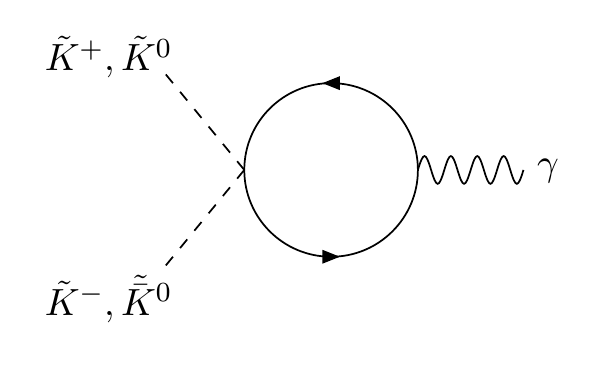}
\caption{ \label{kk-g}
\footnotesize{
The $\phi\mbox{-}\phi\mbox{-}\gamma$ coupling
(charge radius), which vanishes in the $SU(3)_V$ limit.
The $\tilde{\pi}\mbox{-}\tilde{\pi}\mbox{-}\gamma$
coupling vanishes in the $SU(2)_V$ limit.}}
\end{figure}

\noindent
$\bullet$$\phi\mbox{-}\phi \mbox{-}\gamma$\\
The corresponding one-loop diagram is shown in Fig.~\ref{kk-g},
where the right diagram  in Fig.~\ref{kk-g} yields zero contribution.
\be
& &{\cal L}_{\phi^2\gamma}=
A^{\mu\nu}\left(
G_{K^+K^- \gamma}\partial_\mu\tilde{K}^+\partial_\nu\tilde{K}^-
+G_{K^0 \bar{K}^0 \gamma}\partial_\mu\tilde{K}^0
\partial_\nu\tilde{\bar{K}}^0+
G_{\pi^+\pi^- \gamma}\partial_\mu\tilde{\pi}^+\partial_\nu\tilde{\pi}^-
\right)\nn\\
& &+Z^{\mu\nu}\left(
G_{K^+K^- Z}\partial_\mu\tilde{K}^+\partial_\nu\tilde{K}^-
+G_{K^0 \bar{K}^0 Z}\partial_\mu\tilde{K}^0
\partial_\nu\tilde{\bar{K}}^0+
G_{\pi^+\pi^- Z}\partial_\mu\tilde{\pi}^+\partial_\nu\tilde{\pi}^-
\right)~,\label{Lp2g}
\ee
where $A(Z)_{\mu\nu}=\partial_\mu A(Z)_\nu-\partial_\nu A(Z)_\mu$.
The effective couplings in the large $\Lambda$ limit are
\be
G_{K^+K^- \gamma}
&=&2Z_{\tilde{K}^\pm} n_c e Q
\left(1-\frac{G_D}{8 G^2}\langle\sigma_2 \rangle\right)^2
I_{\phi^2\gamma}(M_3,M_1)~,\label{GKKg}\\
G_{K^0\bar{K}^0 \gamma}
&=&2Z_{\tilde{K}^0} n_c e Q
\left(1-\frac{G_D}{8 G^2}\langle\sigma_1 \rangle\right)^2
I_{\phi^2\gamma}(M_3,M_2)~,\nn\\
G_{\pi^+\pi^- \gamma}&=&2Z_{\tilde{\pi}^\pm} n_c e Q
\left(1-\frac{G_D}{8 G^2}\langle\sigma_3 \rangle\right)^2
I_{\phi^2\gamma}(M_2,M_1)~,\nn\\
G_{K^+K^- Z}&=&-t_W G_{K^+K^- \gamma}~,
G_{K^0 \bar{K}^0 Z}=-t_W G_{K^0 \bar{K}^0\gamma}~,
G_{\pi^+\pi^- Z}=-t_W G_{\pi^+\pi^- \gamma}~,\nn
\ee
$t_W^2 =(\sin\theta_W/\cos\theta_W)^2\simeq 0.3$,
and $I_{\phi\gamma^2}(m_a,m_b)$ is given in (\ref{Ippg}).
In the $SU(2)_V$ limit, 
we obtain
$G_{K^0\bar{K}^0 \gamma}
=G_{K^+K^- \gamma}$ and $G_{\pi^+\pi^- \gamma}=0$, because
$I_{\phi^2\gamma}(m_a,m_b)
\to  (m_b-m_a)/(48 \pi^2 m_a^3)$ as  $m_b\to  m_a$.

\begin{figure}
\includegraphics[width=0.6\textwidth]{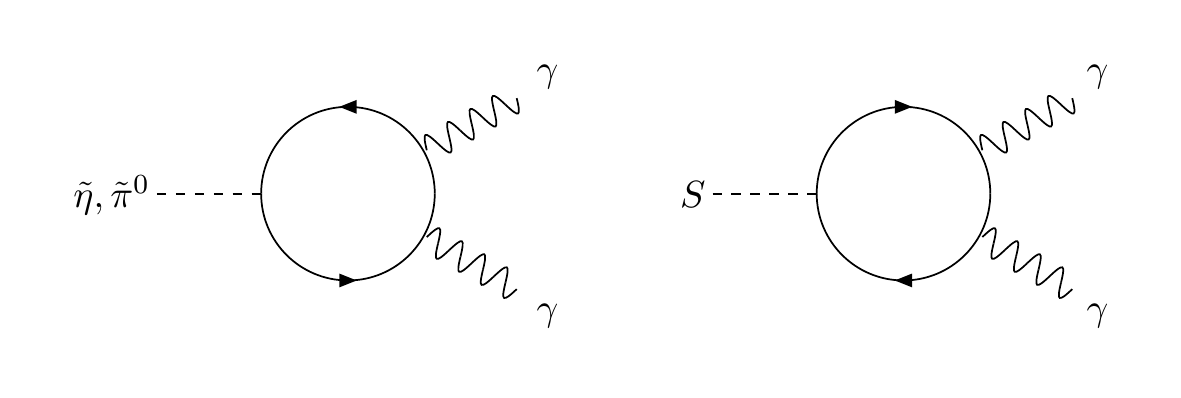}
\caption{ \label{eta-gg}
\footnotesize{Decay of DM and $S$ into two $\gamma$s.
In the $SU(2)_V$ limit  $\tilde{\pi}^0$ does not decay. }}
\end{figure}

\noindent
$\bullet$$\phi
\to \gamma\gamma$\\
The diagram in Fig.~\ref{eta-gg} shows the decay
 of $\tilde{\eta}, \tilde{\pi}^0$ and $S$ into two $\gamma$s, but they can also
decay two $Z$s and  $\gamma$ and $Z$, if the processes are kinematically allowed.
Using the NJL Lagrangian (\ref{L0}) and (\ref{Ipgg}) in Appendix C we find that the effective interaction takes the form
\be
& &{\cal L}_{\phi\gamma^2}=
\frac{1}{4}\tilde{\eta}\epsilon^{\mu\nu\alpha\beta}
\left(
\frac{1}{2}G_{\eta \gamma^2}~A_{\mu\nu}A_{\alpha\beta}+
G_{\eta\gamma Z}~A_{\mu\nu}Z_{\alpha\beta}+
\frac{1}{2}G_{\eta Z ^2}~Z_{\mu\nu}Z_{\alpha\beta}\right)\nn\\
& &+\frac{1}{4}\tilde{\pi}^0\epsilon^{\mu\nu\alpha\beta}
\left(
\frac{1}{2}G_{\pi^0 \gamma^2}~ A_{\mu\nu}A_{\alpha\beta}+
G_{\pi^0\gamma Z}~A_{\mu\nu}Z_{\alpha\beta}
+\frac{1}{2}G_{\pi^0 Z^2}~Z_{\mu\nu}Z_{\alpha\beta}\right)~,
\label{Lpg2}
\ee
where in the large $\Lambda$ limit
\be
G_{\eta \gamma^2}
&=&Z_{\tilde{\eta}}^{1/2} n_c   \frac{\alpha}{\sqrt{3}\pi}Q^2
\left[\left(1-\frac{G_D}{8 G^2}(2\langle\sigma_2 \rangle-\langle\sigma_3 \rangle)\right)M_1^{-1}\right.\nn\\
& &\left.+\left(1-\frac{G_D}{8 G^2}(2\langle\sigma_1 \rangle-\langle\sigma_3 \rangle)\right)M_2^{-1}
-\left(2-\frac{G_D}{8 G^2}(2\langle\sigma_1 \rangle-\langle\sigma_3 \rangle)\right)M_3^{-1}
\right],\label{Geg2}\\
G_{\eta\gamma Z}
&=&- t_W G_{\eta \gamma^2},~
G_{\eta Z^2}
= t_W^2 G_{\eta \gamma^2}~,\nn\\
G_{\pi^0  \gamma^2}
&=&Z_{\tilde{\pi}}^{1/2} n_c  \frac{\alpha}{\pi}Q^2
\left(1-\frac{G_D}{8 G^2}\langle\sigma_3 \rangle\right)\left(M_1^{-1}-
M_2^{-1}\right)~,\label{Gpg2}\\
G_{\pi^0\gamma Z}
&=&- t_W G_{\pi^0\gamma^2},~
G_{\pi^0 Z^2}
= t_W^2 G_{\pi^0 \gamma^2}~.\nn
\ee
As we see from (\ref{Gpg2}), the $\tilde{\pi}^0 \to \gamma\gamma$ decay 
vanishes in the  $SU(2)_V$ limit,
because $M_1=M_2$ in this limit.

The decay of $S$ into two $\gamma$, two $Z$ and $\gamma Z$ can be described
by
\be
& &{\cal L}_{S\gamma^2}=
S\left(
\frac{1}{2}G_{S \gamma^2}~A_{\mu\nu}A^{\mu\nu}+
G_{S\gamma Z}~A_{\mu\nu}Z^{\mu\nu}+
\frac{1}{2}G_{S Z ^2}~Z_{\mu\nu}Z^{\mu\nu}\right)~,
\label{LSg2}
\ee
where we find from (\ref{Sgg})
\be
G_{S \gamma^2}
&=&  \frac{\alpha}{3\pi}Q^2\sum_{i=1,2,3} y_i M_i^{-1},\\
G_{S\gamma Z}
&=&- t_W G_{S \gamma^2},~
G_{S Z^2}
= t_W^2 G_{S \gamma^2}~.
\label{GSg2}~
\ee

\begin{figure}
\includegraphics[width=0.3\textwidth]{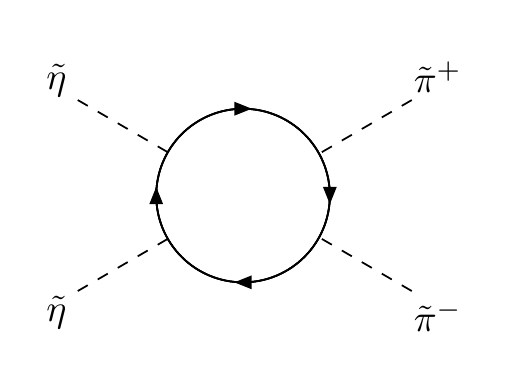}
\includegraphics[width=0.29\textwidth]{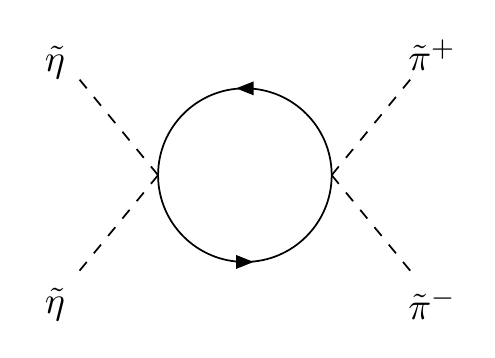}
\caption{ \label{conversion1}
\footnotesize{Examples for DM conversion}}
\end{figure}

\begin{figure}
\includegraphics[width=0.3\textwidth]{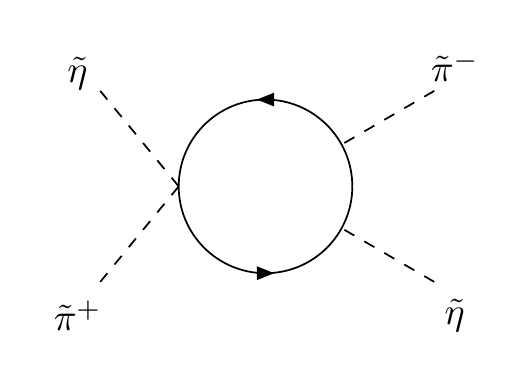}
\caption{ \label{conversion3}
\footnotesize{This DM conversion vanishes in the
$SU(3)_V$ limit.}}
\end{figure}

\noindent
$\bullet$Dark matter conversion\\
The diagrams in Figs.~\ref{conversion1} and \ref{conversion3}
are examples of DM conversion,
in which two incoming DM particles are annihilated into a pair of
two  DM particles which are different from the incoming ones.
There are  DM  conversion amplitudes, which do not 
vanish in the $SU(3)_V$ limit, and those which vanish in the limit.
Except the last
$\eta K^2\pi$ interaction term, the effective interaction term below
do  not vanish  the $SU(3)_V$ limit.
\be
{\cal L}_{\phi^4} &=&\frac{1}{2}G_{\eta^2K^2}~\tilde{\eta}^2 
\left(\tilde{K}^0\tilde{\bar{K}}^0+
\tilde{K}^+\tilde{K}^-\right) +
\frac{1}{2}G_{\eta^2\pi^2}~\tilde{\eta}^2
 \left(\frac{1}{2}(\tilde{\pi}^0)^2+
\tilde{\pi}^+\tilde{\pi}^-\right)
\nn\\
& &+
G_{K^2\pi^2}
\left(\tilde{K}^0\tilde{\bar{K}}^0
+\tilde{K}^+\tilde{K}^-\right)
\left(
\frac{1}{2}(\tilde{\pi}^0)^2+\tilde{\pi}^+\tilde{\pi}^-\right)\nn\\
& &+
G_{\eta K^2\pi}\tilde{\eta}\left(
(\tilde{K}^0\tilde{\bar{K}}^0 
-\tilde{K}^+\tilde{K}^-)\tilde{\pi}^0
+\sqrt{2}\tilde{K}^0\tilde{K}^- \tilde{\pi}^+
+\sqrt{2}\tilde{\bar{K}}^0\tilde{K}^+ \tilde{\pi}^-
\right)~,\label{Lp4}
\ee
where
\be
G_{\eta^2 K^2} &=&
\frac{4}{3}Z_{\tilde{\eta}}Z_{\tilde{K}}n_c
\left(1-\frac{G_D}{8 G^2}\langle \sigma_1\rangle\right)^2
\left[
\left(1-\frac{G_D}{8 G^2}(2\langle \sigma_1\rangle-
\langle \sigma_3\rangle)\right)^2~
 I_{\phi^4}^{2A}(M_1,M_3)\right.\nn\\
 & & +4\left(1-\frac{G_D}{8 G^2}\langle \sigma_1\rangle\right)^2~
 I_{\phi^4}^{2A}(M_3,M_1)\nn\\
 & &\left.-2\left(1-\frac{G_D}{8 G^2}\langle
  \sigma_1\rangle\right)\left(1-\frac{G_D}{8 G^2}(2\langle \sigma_1\rangle-
\langle \sigma_3\rangle)\right)~ I_{\phi^4}^{1A}(M_1,M_3)\right]\nn\\
& &+\frac{4}{3}Z_{\tilde{\eta}}Z_{\tilde{K}}n_c
\left( \frac{G_D}{8 G^2}\right)^2\left(
 I_{\phi^4}^{1B}(M_1,M_3)+
 2 I_{\phi^4}^{2B}(M_1)\right)~,
\label{Ge2K2}\\
G_{\eta^2\pi^2} &=&
4 Z_{\tilde{\eta}}Z_{\tilde{\pi}}n_c
\left(1-\frac{G_D}{8 G^2}\langle \sigma_3\rangle\right)^2
\left(1-\frac{G_D}{8 G^2}(2\langle \sigma_1\rangle-
\langle \sigma_3\rangle)\right)^2 
 I_{\phi^4}^{3A}(M_1)\nn\\
& &+\frac{4}{3}Z_{\tilde{\eta}}Z_{\tilde{\pi}}n_c
\left( \frac{G_D}{8 G^2}\right)^2\left(
4  I_{\phi^4}^{2B}(M_1)-
 I_{\phi^4}^{2B}(M_3)\right)~,\label{Ge2p2}\\
G_{K^2\pi^2} &=&
4 Z_{\tilde{K}}Z_{\tilde{\pi}}n_c
\left(1-\frac{G_D}{8 G^2}\langle \sigma_1\rangle\right)^2
\left(1-\frac{G_D}{8 G^2}\langle \sigma_3\rangle\right)^2 
 I_{\phi^4}^{2A}(M_1,M_3)\nn\\
& &+4Z_{\tilde{\pi}}Z_{\tilde{K}}n_c
\left( \frac{G_D}{8 G^2}\right)^2
 I_{\phi^4}^{1B}(M_1,M_3)~,\label{K2p2}\\
G_{\eta K ^2\pi} &=&
\frac{4}{\sqrt{3}}Z_{\tilde{\eta}}^{1/2}Z_{\tilde{K}}
Z_{\tilde{\pi}}^{1/2}n_c
\left(1-\frac{G_D}{8 G^2}\langle \sigma_1\rangle\right)^2
\left(1-\frac{G_D}{8 G^2}\langle \sigma_3\rangle\right)\nn\\ 
& &\times\left[
\left(1-\frac{G_D}{8 G^2}\langle \sigma_1\rangle\right)~
 I_{\phi^4}^{1A}(M_1,M_3)
-\left(1-
\frac{G_D}{8 G^2} (2\langle \sigma_1\rangle-\langle \sigma_3\rangle)\right)~
 I_{\phi^4}^{2A}(M_1,M_3)\right]\nn\\
& &+\frac{4}{\sqrt{3}}Z_{\tilde{\eta}}^{1/2}Z_{\tilde{K}}
Z_{\tilde{\pi}}^{1/2}n_c
\left( \frac{G_D}{8 G^2}\right)^2\left(
 I_{\phi^4}^{1B}(M_1,M_3)-
  I_{\phi^4}^{2B}(M_1)\right)~,\label{GeK2p}
\ee
and $I_{\phi^4}^{1A}(m_a,m_b)$ etc. are defined in 
(\ref{Ip41A})-(\ref{Ip42B})  in Appendix C.
We have not included the contributions from the diagram
like one in Fig.~\ref{conversion3},
because they are negligibly suppressed in a 
realistic parameter space, in which $SU(3)_V$ is only weakly broken. 
Similarly, $G_{\eta K ^2\pi}$, too, is negligibly small
($G_{\eta K^2 \pi}/G_{\eta^2 K^2} \sim 10^{-4}$), so that we will not take into account
the $\eta K^2\pi$ interactions in computing the DM  relic abundance.

\begin{figure}
\includegraphics[width=0.3\textwidth]{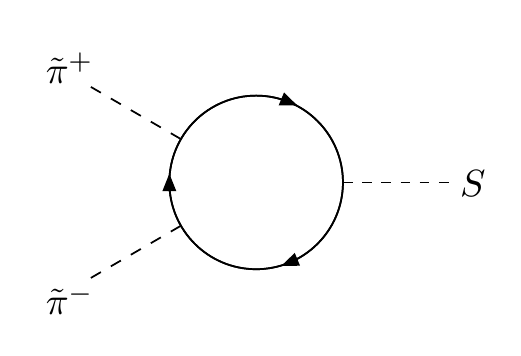}
\includegraphics[width=0.3\textwidth]{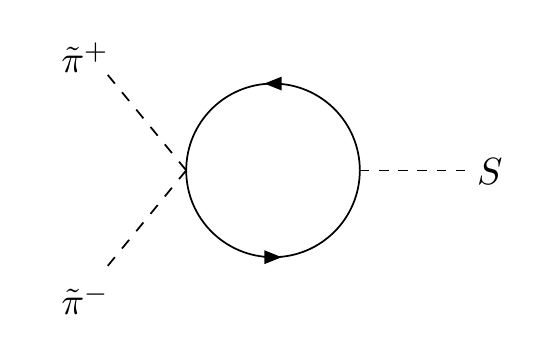}
\caption{ \label{pp-s}
\footnotesize{DM coupling with one $S$. In a realistic 
parameter space there is an accidental cancellation
between these two diagrams.}}
\end{figure}

\begin{figure}
\includegraphics[width=0.31\textwidth]{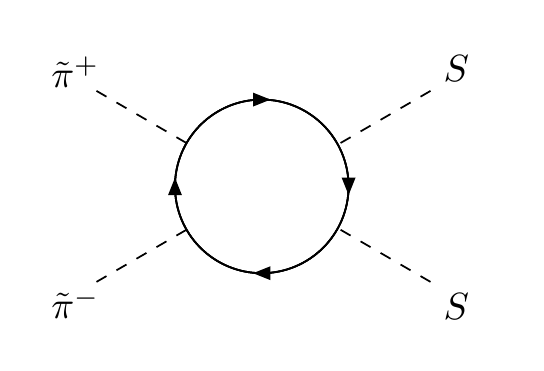}
\includegraphics[width=0.28\textwidth]{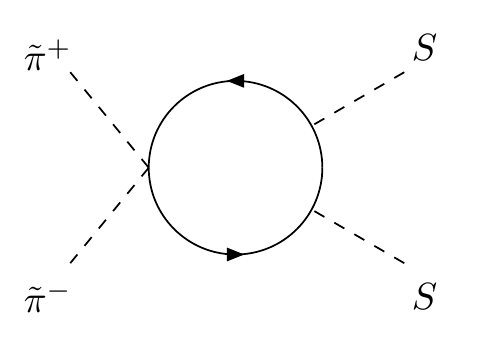}
\caption{ \label{pp-ss}
\footnotesize{DM coupling with two $S$s.
These diagrams contribute to the DM relic abundance
if the mass of $S$ is comparable with or
less than the DM masses.
}}
\end{figure}

\noindent
$\bullet$Dark matter coupling with $S$\\
The diagrams in Figs.~\ref{pp-s} and \ref{pp-ss} show
dark matter interactions with the singlet $S$.
The DM  coupling with $S$ (Fig.~\ref{pp-s}) can be described by
\be
{\cal L}_{\phi^2 S} &=& 
S\left(
\frac{1}{2}G_{\eta^2 S}  \tilde{\eta}^2+
+G_{K^2 S} (\tilde{K}^0\tilde{\bar{K}}^0
+\tilde{K}^+\tilde{K}^-)+
G_{\pi^2 S} (\frac{1}{2}(\tilde{\pi}^0)^2
+\tilde{\pi}^+\tilde{\pi}^-)
\right)~.\label{Lp2S}
\ee
Using (\ref{Ip2s1A}) - (\ref{Ip2sB}) in Appendix C we find in the large $\Lambda$ limit
\be
G_{\eta^2 S} &=&
-\frac{2}{3} Z_{\tilde{\eta}}n_c
\left[
4 y_1\left(1-\frac{G_D}{8G^2}
(2\langle\sigma_1\rangle-\langle\sigma_3\rangle)
\right)^2 I^{2A}_{\phi^2 S}(M_1)+
2 y_3\left(1-\frac{G_D}{8G^2}
\langle\sigma_1\rangle\right)^2 I^{2A}_{\phi^2 S}(M_3)\right]\nn\\
& &
-\frac{2}{3} Z_{\tilde{\eta}}n_c\left(\frac{G_D}{8G^2}\right)
\left(4 y_1 I^{B}_{\phi^2 S}(M_1)
-y_3 I^{B}_{\phi^2 S}(M_3)
\right)~,\label{Ge2S}\\
G_{K^2 S} &=&
 -2  Z_{\tilde{K}}n_c\left(1-\frac{G_D}{8G^2}
\langle\sigma_1\rangle\right)^2
\left(
y_1 I^{1A}_{\phi^2 S}(M_3,M_1)+
y_3 I^{1A}_{\phi^2 S}(M_1,M_3)\right)\nn\\
& &
-2  Z_{\tilde{K}}n_c\left(\frac{G_D}{8G^2}\right)y_1 I^{B}_{\phi^2 S}(M_1)~,
\label{GK2S}\\
G_{\pi^2 S} &=&
-4  Z_{\tilde{\pi}}n_c\left(1-\frac{G_D}{8G^2}
\langle\sigma_3\rangle\right)^2y_1 I^{2A}_{\phi^2 S}(M_1)-
2  Z_{\tilde{\pi}}n_c\left(\frac{G_D}{8G^2}\right)y_3I^{B}_{\phi^2 S}(M_3)~.\label{Gp2S}
\ee

The DM  coupling with two $S$s (Fig.~\ref{pp-ss}) can be described by
\be
{\cal L}_{\phi^2 S^2}&= &\frac{1}{2} S^2\left(
\frac{1}{2}G_{\eta^2 S^2}  \tilde{\eta}^2+
+G_{K^2 S^2} (\tilde{K}^0\tilde{\bar{K}}^0
+\tilde{K}^+\tilde{K}^-)+
G_{\pi^2 S^2} (\frac{1}{2}(\tilde{\pi}^0)^2
+\tilde{\pi}^+\tilde{\pi}^-)
\right)~,\label{Lp2S2}
\ee
where
\be
G_{\eta^2 S^2} &=&
-\frac{2}{3}  Z_{\tilde{\eta}}n_c
\left[
y_1^2 \left(1-\frac{G_D}{8G^2}
(2\langle\sigma_1\rangle-\langle\sigma_3\rangle)\right)^2 
\left(2 I_{\phi^2S^2}^{2A}(M_1)
+ I_{\phi^2 S^2}^{2B}(M_1)\right) \right.\nn\\
& &\left.+2 y_3^2 \left(1-\frac{G_D}{8G^2}
\langle\sigma_1\rangle\right)^2 
\left(2 I_{\phi^2S^2}^{2A}(M_3)
+ I_{\phi^2 S^2}^{2B}(M_3)\right) \right]\nn\\
& &- \frac{1}{3} Z_{\tilde{\eta}}n_c\left(\frac{G_D}{8G^2}\right)
\left(
4 y_1^2~ I^{C}_{\phi^2 S^2}(M_1)-
 y_3^2~ I^{C}_{\phi^2 S^2}(M_3)\right)~,\label{Ge2S2}\\
G_{K^2 S^2} &=&
-2  Z_{\tilde{K}}n_c\left(1-\frac{G_D}{8G^2}
\langle\sigma_1\rangle\right)^2 
\left(
y_1^2 ~ I^{1A}_{\phi^2 S^2}(M_3,M_1)+ 
y_3^2 ~ I^{1A}_{\phi^2 S^2}(M_1,M_3)+ 
y_1 y_3 I^{1B}_{\phi^2 S^2}(M_1,M_3)  \right)\nn\\
& &- Z_{\tilde{K}}n_c\left(\frac{G_D}{8G^2}\right)
y_1^2~ I^{C}_{\phi^2 S^2}(M_1)~,\label{GK2S2}\\
 G_{\pi^2 S^2} &=&
-2  Z_{\tilde{\pi}}n_c\left(1-\frac{G_D}{8G^2}
\langle\sigma_3\rangle\right)^2 y_1^2 
\left(
2 I^{2A}_{\phi^2 S^2}(M_1)+ I^{2B}_{\phi^2 S^2}(M_1)  \right)\nn\\
& &-  Z_{\tilde{\pi}}n_c\left(\frac{G_D}{8G^2}\right)
y_3^2 I^{C}_{\phi^2 S^2}(M_3)~.\label{Gp2S2}
\ee

\begin{figure}
\includegraphics[width=0.3\textwidth]{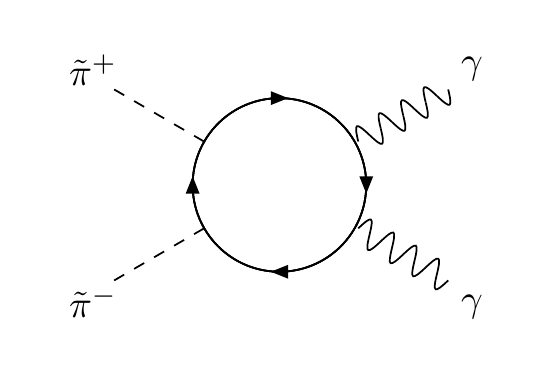}
\includegraphics[width=0.27\textwidth]{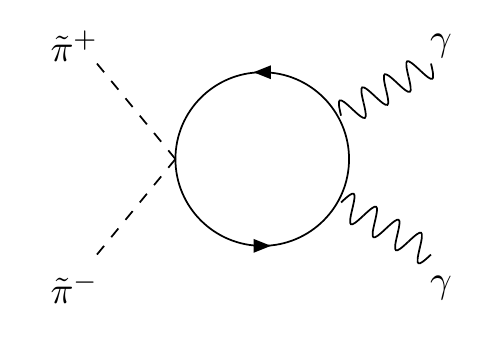}
\caption{ \label{pp-to-gg}
\footnotesize{DM annihilation into two $\gamma$s.}}
\end{figure}

\noindent
$\bullet$Dark matter coupling with two $\gamma$s\\
The diagram in Fig.~\ref{pp-to-gg} shows
 the annihilation of $\pi^\pm$ pair into
two $\gamma$s, where   the annihilations into 
$\gamma Z$, two $Z$s and also into two $S$s
are also possible if they are kinematically allowed.
\be
{\cal L}_{\phi^2G^2} &=& 
\frac{1}{4}A_{\mu\nu} A^{\mu\nu}\left(
G_{\eta^2\gamma^2}\frac{1}{2}(\tilde{\eta})^2+
G_{K^2\gamma^2} (\tilde{K}^0\tilde{\bar{K}}^0
+\tilde{K}^+\tilde{K}^- )+
G_{\pi^2\gamma^2} (\frac{1}{2}(\tilde{\pi}^0)^2
+\tilde{\pi}^+\tilde{\pi}^-)
\right)\nn\\
& &+
\frac{1}{2}A_{\mu\nu} Z^{\mu\nu}\left(
G_{\eta^2\gamma Z}\frac{1}{2}(\tilde{\eta})^2
+
G_{K^2\gamma Z} (\tilde{K}^0\tilde{\bar{K}}^0
+\tilde{K}^+\tilde{K}^-)+
G_{\pi^2\gamma Z} (\frac{1}{2}(\tilde{\pi}^0)^2
+\tilde{\pi}^+\tilde{\pi}^-)
\right)\label{Lp2G2}\\
& &+
\frac{1}{4}Z_{\mu\nu} Z^{\mu\nu}\left(
G_{\eta^2 Z^2}\frac{1}{2}(\tilde{\eta})^2
+
G_{K^2 Z^2} (\tilde{K}^0\tilde{\bar{K}}^0
+\tilde{K}^+\tilde{K}^-)+
G_{\pi^2 Z^2} (\frac{1}{2}(\tilde{\pi}^0)^2
+\tilde{\pi}^+\tilde{\pi}^-)
\right),\label{p2G2}\nn
\ee
where
$A(Z)_{\mu\nu}=\partial_\mu A(Z)_\nu-\partial_\nu A(Z)_\mu$.
Using the approximate form (\ref{Ip2g2A}) and (\ref{Ip2g2C}) we find
\be
G_{\eta^2\gamma^2} &=&Z_{\tilde{\eta}} n_c
\frac{\alpha }{ \pi}Q^2 {\cal A}_{\tilde{\eta}}
(\gamma\gamma)\simeq Z_{\tilde{\eta}}n_c
\frac{\alpha }{ \pi}Q^2 ~
{\cal A}_{\tilde{\pi}}(\gamma\gamma)~,
\label{e2g2}\\
G_{\eta^2\gamma Z} &=& -t_W G_{\eta^2\gamma^2},~
G_{\eta^2 Z^2}=t_W^2 G_{\eta^2\gamma^2},\nn\\
G_{K^2\gamma^2} &=&Z_{\tilde{K}}n_c\frac{\alpha }{ \pi}Q^2 ~
{\cal A}_{\tilde{K}}(\gamma\gamma)
\simeq Z_{\tilde{K}}n_c\frac{\alpha }{ \pi}Q^2 ~
{\cal A}_{\tilde{\pi}}(\gamma\gamma)~,\nn\\
G_{K^2\gamma Z}&=& -t_W G_{K^2\gamma^2},~
G_{K^2 Z^2}=t_W^2 G_{K^2\gamma^2},\nn\\
G_{\pi^2\gamma^2} &=& Z_{\tilde{\pi}}n_c\frac{\alpha }{ \pi}Q^2 ~
{\cal A}_{\tilde{\pi}}(\gamma\gamma)~,
\label{p2g2}\\
G_{\pi^2\gamma Z}&=& -t_W G_{\pi^2\gamma^2},~
G_{\pi^2 Z^2}=t_W^2 G_{\pi^2\gamma^2},\nn
\ee
where
\be
{\cal A}_{\tilde{\pi}}(\gamma\gamma)
 &= & \frac{4}{3}
 \left[ - \left(1-\frac{G_D}{8 G^2}\langle\sigma_3\rangle
 \right)^2 M_1^{-2}+   
 \frac{G_D}{8 G^2}M_3^{-1}\right]~,\label{Agg}
\ee
and 
${\cal A}_{\tilde{\eta}}(\gamma\gamma)=
{\cal A}_{\tilde{K}}(\gamma\gamma)=
{\cal A}_{\tilde{\pi}}(\gamma\gamma)$
in the $SU(3)_V$ limit. In a realistic  parameter space
for the $SU(2)_V$ case, the ratio
${\cal A}_{\tilde{\pi}}(\gamma\gamma)/ 
{\cal A}_{\tilde{\eta}}(\gamma\gamma)$, for instance, is at most $1.004$.

In the following discussions we shall use the effective interaction terms 
derived above  to compute the DM relic abundance as well as
the cross sections for the direct and indirect detections of DM.

\subsection{Relic Abundance of Dark Matter}
The $SU(3)_V$ case (\ref{su3}) has been discussed in 
\cite{Holthausen:2013ota,Kubo:2014ida}, and
so we below consider only the (i) and (ii) cases,
which are defined in (\ref{u1u1}) and (\ref{su2u1}), respectively.
In a one-component DM system, the velocity-averaged annihilation cross section
$\langle v \sigma \rangle$ should be $\sim 10^{-9}~\mbox{GeV}^{-2}$ to
obtain a realistic DM relic abundance $\Omega h^2\simeq 0.12$.
A rough estimate of the velocity-averaged annihilation cross section
for DM conversion (Fig.~\ref{conversion1} ) shows
$\langle v \sigma (\tilde{\eta}\tilde{\eta}
 \to \tilde{\pi}^+ \tilde{\pi}^-) \rangle
\simeq 10^{-5} (1-m_{\tilde{\pi}}^2/m_{\tilde{\eta}}^2)^{1/2} 
~\mbox{GeV}^{-2}$, where it vanishes if $SU(3)_V$ is unbroken.
The reason for the large  annihilation cross section  for  DM conversion 
is that the coupling of the hidden fermions to the
hidden mesons is of $O(1)$: There is no coupling constant for the coupling
as one can see from the NJL Lagrangian (\ref{L0}).
That is, the annihilation cross section  for  DM conversion is about
four orders of magnitude larger than that in an ordinary case, unless
the masses of the incoming and outgoing DMs are almost  degenerate.

\subsubsection{(i) $U(1)_{\tilde{B'}}\times U(1)_{\tilde{B}}$ }

There exists a problem for the $U(1)_{\tilde{B'}}\times U(1)_{\tilde{B}}$ 
case (\ref{u1u1}), which we will discuss now.
As we have found in the previous subsection, 
the lightest NG boson in the $U(1)_{\tilde{B'}}\times U(1)_{\tilde{B}}$ case
is  always the lightest between the neutral ones within the one-loop analysis
in the NJL approximation and that without of loss of generality we can assume
it is $\tilde{\pi}^0$. Its dominant decay mode is into two $\gamma$s.
The decay width  can be calculated from the effective
 Lagrangian (\ref{Lpg2}):
\be
\Gamma(\tilde{\pi}^0\to \gamma\gamma)
&=&\frac{9Z_{\tilde{\pi}} Q^4\alpha^2}{64\pi^3} m_{\tilde{\pi}^0}^3
\left(1-\frac{G_D}{8G^2}\langle\sigma_3\rangle\right)^2\left(
1/M_1 -1/M_2\right)^2
\nn\\
& \simeq & 10^{-6}\times Q^4 m_{\tilde{\pi}^0}^3\left(
1/M_1 -1/M_2\right)^2~,\label{ptogg}
\ee
which should be compared with the expansion rate $H$ of the Universe
at $T=m_{\tilde{\pi}^0}$,
\be
\frac{\Gamma(\tilde{\pi}^0\to \gamma\gamma)}{H}
& \simeq & 7\times 10^{8}~Q^4 ( m_{\tilde{\pi}^0}/M_1)^3
(\Delta M/M_1)^2 \left[\mbox{TeV}/M_1\right]+O(\Delta M^3)~,
\ee
where $\Delta M=M_1-M_2$.
Therefore, unless the $U(1)_Y$ charge $Q$ of the hidden 
fermions is very small or the constituent fermion masses $M_1$ and $M_2$ are 
accurately fine-tuned (or both), 
$\tilde{\pi}^0$ decays immediately into two $\gamma$s.
Since the stable DM particles can annihilate into
two $\tilde{\pi}^0$s with a huge DM conversion rate,
there will be almost no DM left in the end.
Since  we want to assume 
neither a tinny $Q$ nor accurately fine-tuned  constituent fermion masses,
we will not consider below the DM phenomenology
based on the $U(1)_{\tilde{B'}}\times U(1)_{\tilde{B}}$ flavor symmetry.

\subsubsection{(ii) $SU(2)_V \times U(1)_{\tilde{B}}$ }
Now we come to the case (ii) in (\ref{su2u1}),
which means $y_1=y_2 < y_3$. In this case the unstable NG boson is
$\tilde{\eta}$ which can decay into two $\gamma$s (and also into two
$S$s if it is kinematically allowed). Because of $SU(2)_V$, $\tilde{\pi}^0$ 
is now stable and $m_{\tilde{\pi}}=m_{\tilde{\pi}^0}=m_{\tilde{\pi}^\pm}$
(see (\ref{hrcy})).
Furthermore, $m_{\tilde{K}^0}$ and $m_{\tilde{K}^\pm}$,
which are slightly larger than  $m_{\tilde{\pi}}$, are exactly degenerate
in this case, i.e. $m_{\tilde{K}}=m_{\tilde{K}^\pm}= m_{\tilde{K}^0}$.
In the parameter region (\ref{y-range}) we can further constrain the parameter space.
Since $y$ is a measure of the explicit chiral symmetry breaking
and at the same time is  the strength of the connection to the SM side,
the smaller is $y$, the smaller is the DM mass, and the larger is the cutoff
$\Lambda$. We have also found that for a given set of
$\lambda_{H},\lambda_{HS}$ and $\lambda_{S}$ the value $ \langle S \rangle$ remains
approximately  constant as $y$ varies, implying that
$m_S$ also remains approximately constant 
because the Higgs mass $m_h\simeq 126$ GeV
and $v_h=\langle h \rangle\simeq 246$ GeV
have to have a fixed value  whatever $y$  is. Consequently, the DM masses are smaller  than
$m_S$, unless  $y_i\gsim 0.015$ or $\lambda_S$ and $\lambda_{HS}$
are very small (or both). We have found,   as long as we assume (\ref{y-range}),
that $\lambda_S \lsim 0.03$ and $\lambda_{HS} \lsim 0.04$ have to be satisfied to realize that $S$ is lighter than DM. However, these values of
$\lambda_S$ and $\lambda_{HS}$  are too small
for the DM annihilation cross sections into two $S$s 
(diagrams in Fig.~\ref{pp-ss} ) to make 
the DM relic abundance realistic.
In summary, there are three groups of DM in the $SU(2)_V$ case (\ref{su2u1});
the heaviest decaying $SU(2)_V$ singlet $\tilde{\eta}$, 
two $SU(2)_V$ doublet
$(\{ \tilde{\bar{K}}^0, \tilde{K}^-\},
 \{\tilde{K}^+,\tilde{K}^0\} )$ and  lightest $SU(2)_V$ triplet 
  $(\tilde{\pi}^\pm, \tilde{\pi}^0)$.
\begin{figure}
\includegraphics[width=0.35\textwidth]{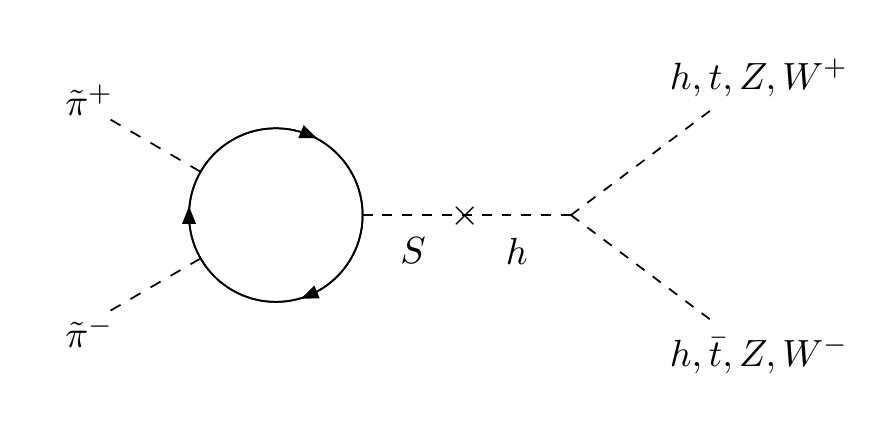}
\includegraphics[width=0.33\textwidth]{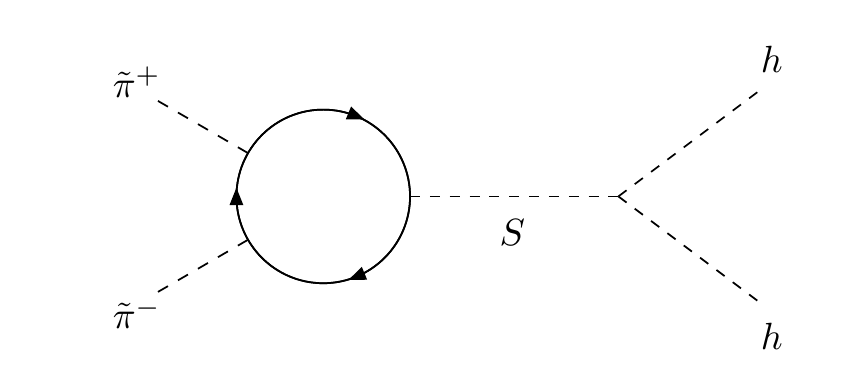}
\caption{ \label{pp-to-SMSM}
\footnotesize{DM annihilation into the SM particles
via an internal $S$ line and $S-h$ mixing.
In the actual calculation of the cross section we use the localized 
expression for the one-loop part, i.e.
$G_{\eta^2S}$ etc given in (\ref{Ge2S})-(\ref{Gp2S}).
}}
\end{figure}

Before we compute the DM relic abundance, let us  simplify
the DM notion:
\be
 \chi_1 &=&\tilde{\eta},~\chi_2~\mbox{to represent}~
\tilde{\bar{K}}^0, \tilde{K}^\pm, \tilde{K}^0,~
 \chi_3~\mbox{to represent}~\tilde{\pi}^\pm, \tilde{\pi}^0
 \ee
with the masses
\be
m_1&=& m_{\tilde{\eta}},~m_2=
 m_{\tilde{K}} ~\mbox{and}~ m_3 = m_{\tilde{\pi}}~(m_1 > m_2 >m_3)~,
 \ee
respectively,
where $\chi_i$ are real scalar fields.
There are  three types of annihilation processes
which enter into the Boltzmann equation:
\be
& &\chi_i~\chi_i \leftrightarrow X~X~,
\label{p0}\\
 & &  \chi_i~\chi_j  \leftrightarrow \chi_k~\chi_l~,
 \label{p1}
\ee
in addition to the decay of $\chi_1$ into two $\gamma$s,
where $X$ stands for the SM particles, and the second process (\ref{p1})
is called  DM conversion.
There are two types of  diagrams for the annihilation into
the SM particles, Fig.~\ref{pp-to-gg} and Fig.~\ref{pp-to-SMSM}.
The diagrams in Fig.~\ref{pp-to-SMSM}
are examples, in which  a one-loop diagram and a tree-diagram are connected
by an internal $S$ or a $S-h$ mixing.
The same process can be realized by using
the right diagram in Fig.~\ref{pp-s} 
for the one-loop part.
It turns out that there is an accidental cancellation
between these two diagrams
so that the velocity-averaged annihilation cross section
is at most $\sim 10^{-11}~\mbox{GeV}^{-2}$, unless near the
resonance in the s- channel \cite{Holthausen:2013ota}.
The effective $\phi\mbox{-}\phi\mbox{-}\gamma$ interaction
(\ref{Lp2g}) can also contribute to the s-channel annihilation into the SM particles.
However, as we have mentioned, the effective coupling
$G_{K^+K^-\gamma}$
is very small in the realistic parameter space.
 For instance, $m_{\tilde{K}}^3G_{K^+K^-\gamma}/
G_{K^2 S}\sim 10^{-5}$, where $G_{K^+K^-\gamma}$ and $G_{K^2 S}$
are given in (\ref{GKKg}) and (\ref{GK2S}), respectively.
Note also that the DM conversion with three different DMs involved 
is forbidden by $SU(3)_V$.
In the $SU(2)_V\times U(1)_{\tilde{B}}$ case, for instance,
$\tilde{\eta}~\tilde{\pi}^- \to \tilde{K}^0~\tilde{K}^-$ is indeed allowed. 
However, it is strongly suppressed 
($G_{\eta K^2 \pi}/G_{\eta^2 K^2} \sim 10^{-4}$),
because $SU(3)_V$ is only weakly  broken in the realistic parameter space.
So we will ignore this type of processes, too, 
in the Boltzmann equation.

Using the notion for  thermally averaged cross sections and decay width
(of $\chi_1$)
\be
 <\!v\sigma (ii;X X) \!>~,~ <\!v\sigma (ii;jj) \!>~,
 <\!\Gamma (1;\gamma\gamma)\!>~,
 \ee
the reduced mass $1/\mu=(\sum_i m_i^{-1})$
and the inverse temperature $x=\mu/T$,
we find  for the number per comoving volume $Y_i= n_i/s$ \cite{Aoki:2012ub}
  \be
& &\frac{d Y_1}{dx}=
-0.264~ g_*^{1/2} \left[\frac{\mu M_{\rm PL}}{x^2} \right]
\left\{<\!v\sigma (11;X X) \!>
\left(  Y_1 Y _1-\bar{Y}_{1}\bar{Y}_{1}\right)\right.\nn\\
& &\left.
+<\!v\sigma (11;22)\!>\!\!\left(  Y_1 Y _1-\frac{Y_2 Y_2}
{\bar{Y}_{2}\bar{Y}_{2}} \bar{Y}_{1}\bar{Y}_{1}
\right)
+<\!v\sigma (11;33)\!>\!\!\left(  Y_1 Y _1-\frac{Y_3 Y_3}
{\bar{Y}_{3}\bar{Y}_{3}} \bar{Y}_{1}\bar{Y}_{1}
\right)~\right\}\\
 & &-0.602~ g_*^{-1/2} \left[\frac{x M_{\rm PL}}{\mu^2} \right]
~ <\!\Gamma (1;\gamma\gamma)\!>
  (Y_1-\bar{Y}_1)~,\nn\\
& &\frac{d Y_2}{dx}=
-0.264~ g_*^{1/2} \left[\frac{\mu M_{\rm PL}}{x^2} \right]
\left\{<\!v\sigma (22;X X) \!>
\left(  Y_2 Y _2-\bar{Y}_{2}\bar{Y}_{2}\right)\right.\nn\\
& &\left.-<\!v\sigma (22;11)\!>\!\!\left(  Y_1 Y _1-\frac{Y_2 Y_2}
{\bar{Y}_{2}\bar{Y}_{2}} \bar{Y}_{1}\bar{Y}_{1}
\right)
+<\!v\sigma (22;33)\!>\!\!\left(  Y_2 Y _2-\frac{Y_3 Y_3}
{\bar{Y}_{3}\bar{Y}_{3}} \bar{Y}_{2}\bar{Y}_{2}\right)\right\}~,\\
& &\frac{d Y_3}{dx}=
-0.264~ g_*^{1/2} \left[\frac{\mu M_{\rm PL}}{x^2} \right]
\left\{<\!v\sigma (33;X X) \!>
\left(  Y_3 Y _3-\bar{Y}_{3}\bar{Y}_{3}\right)\right.\nn\\
& &\left.-<\!v\sigma (33;11)\!>\!\!\left(  Y_1 Y _1-\frac{Y_3 Y_3}
{\bar{Y}_{3}\bar{Y}_{3}} \bar{Y}_{1}\bar{Y}_{1}
\right)
-<\!v\sigma (33;22)\!>\!\!\left(  Y_2 Y _2-\frac{Y_3 Y_3}
{\bar{Y}_{3}\bar{Y}_{3}} \bar{Y}_{2}\bar{Y}_{2}
\right)~\right\}~,
\label{boltz3}
\ee
where
$g_*=115.75$ is the total number of effective degrees of freedom, 
$s$ is the entropy density,
$M_{\rm PL}$ is the Planck mass, and $\bar{Y}_i=n_i/s$ 
in thermal equilibrium.
Although $\langle v\sigma(ii;XX) \rangle$ is much smaller than $10^{-9}~\mbox{GeV}^{-2}$,
we can obtain a realistic value of $\Omega h^2$.
The mechanism is the following \cite{Tulin:2012uq}.
If $y_3$ does not differ very much from $y_1=y_2$,  the differences
among $m_1, m_2$ and $m_3$ are small.
Then at finite temperature inverse DM conversions
(which are kinematically forbidden at  zero temperature)
can become operative, because the DM conversions cross sections
are large, i.e. $10^{-5}~\mbox{GeV}^{-2}\times$ phase space,
as we have mentioned above.
That is, the inverse conversion $\chi_3\chi_3, \chi_2\chi_2 \to
\chi_1\chi_1 \to \gamma\gamma \gamma\gamma$  can play 
a significant role.

\begin{figure}
\includegraphics[width=0.4\textwidth]{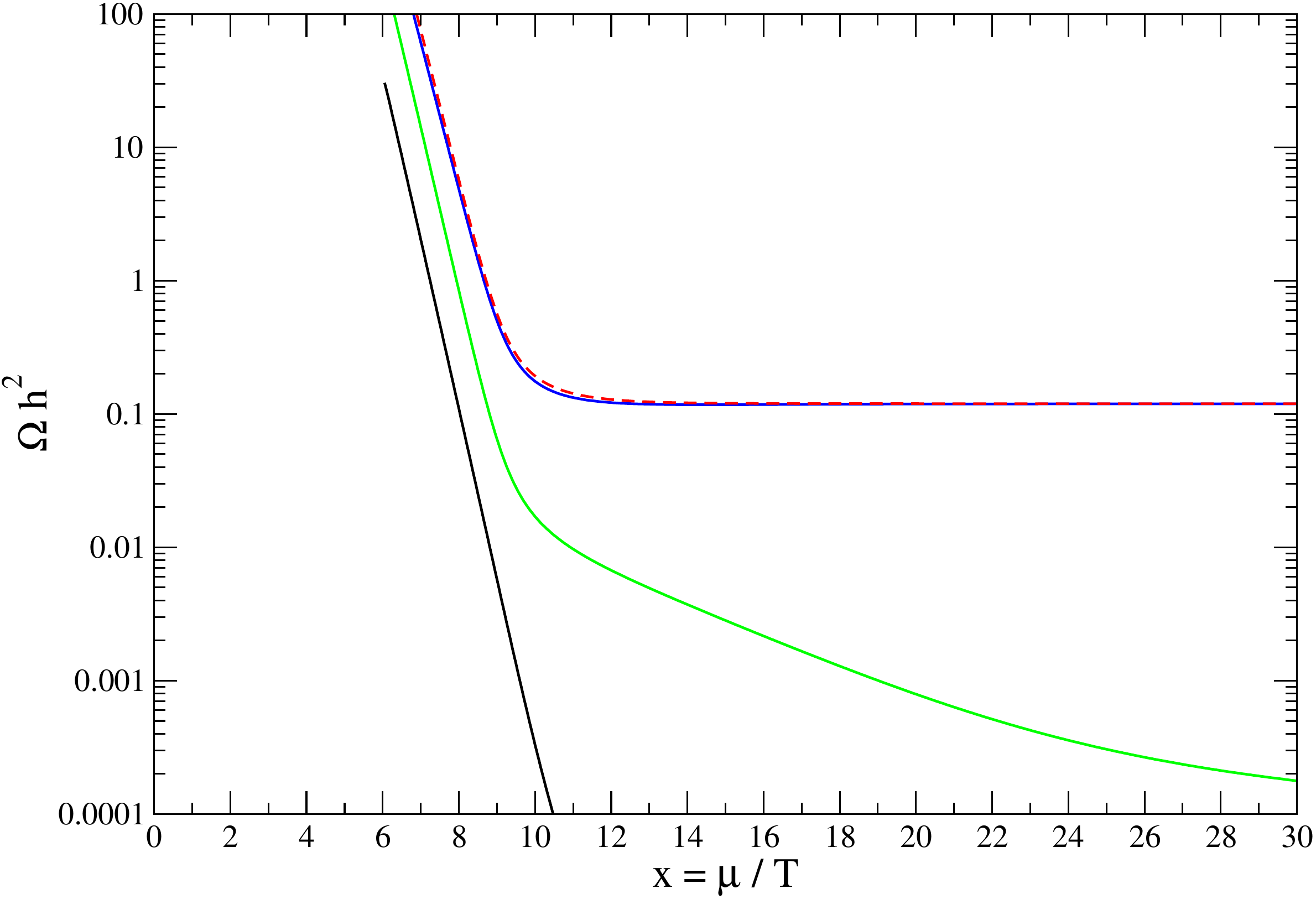}
\includegraphics[width=0.4\textwidth]{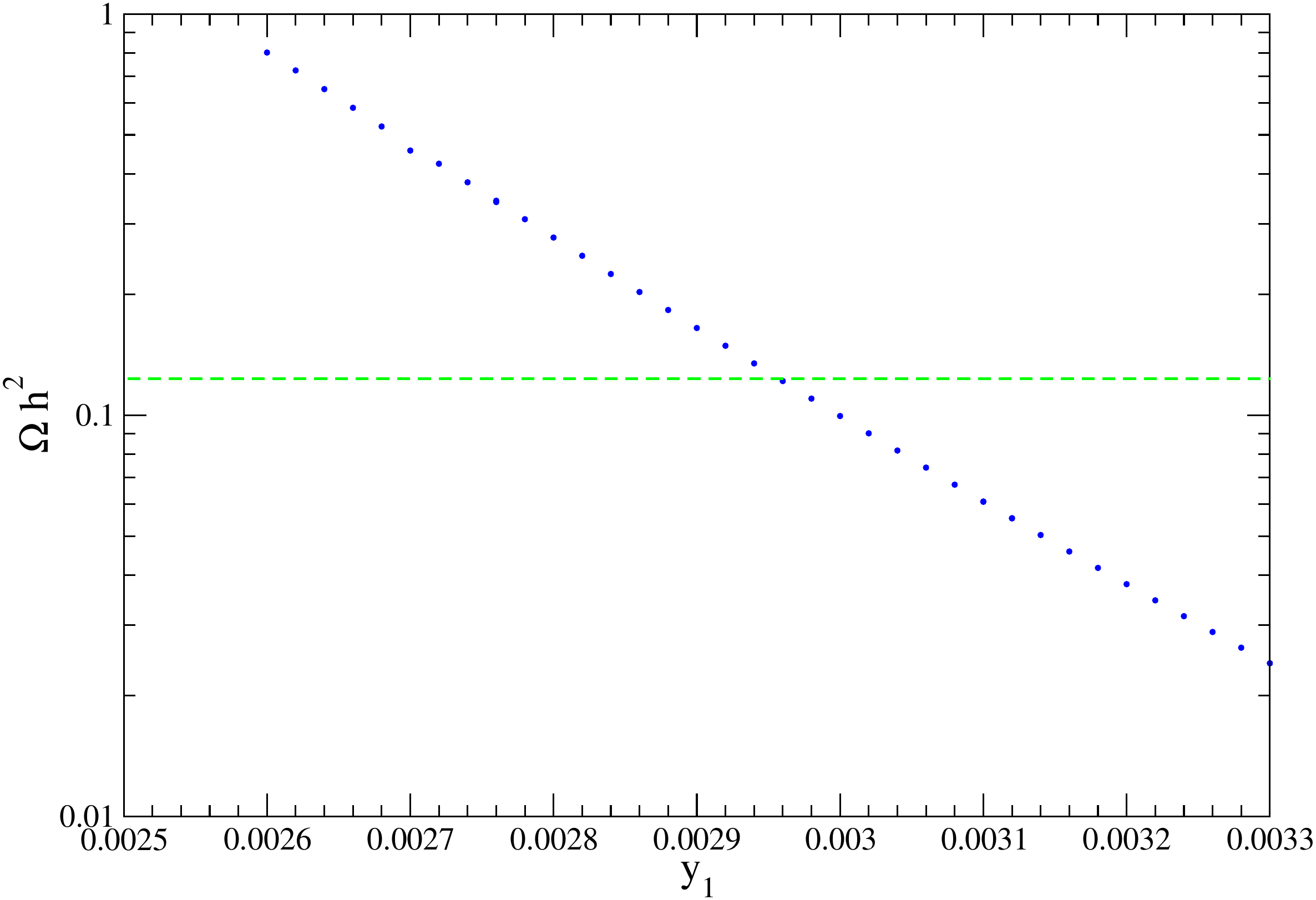}
\caption{ \label{omegas}
\footnotesize{
Left: The relic DM abundances, $\Omega h^2~(\mbox{red}),~
 \Omega_1 h^2=
 \Omega_\eta h^2~(\mbox{black}),~
4\Omega_2 h^2=4\Omega_K h^2 ~(\mbox{green})$ and 
$3\Omega_3 h^2=3\Omega_\pi h^2~(\mbox{blue})$,
 as a function of the inverse temperature $x=\mu/T$
for the parameter values (\ref{parameter}).
Though $\tilde{\eta}$ is almost in thermal equilibrium, its presence is
essential for the $\tilde{K}$ and  $\tilde{\pi}$ numbers to decrease
as $x$ decreases.
In the end  the relic abundance of $\tilde{\pi}$
dominates.
Right: The total relic abundance $\Omega h^2$ as a function
of $y_1(=y_2)$ for $y_3=0.00424$,
where the other parameters are fixed as (\ref{parameter}).}
}
\end{figure}

The relic abundance
$\Omega h^2$ is given by
\be
\Omega_i h^2 &=&
\frac{Y_{i \infty} s_0 m_{i}}{\rho_c/h^2}~,
\label{omega}
\ee
where $Y_{i\infty}$ is the asymptotic value of  $Y_{i}$,
$s_0=2890/\mbox{cm}^3$ is the entropy density at present,
$\rho_c=3 H^2/8 \pi G=
1.05 \times 10^{-5} h^2 ~\mbox{GeV}/\mbox{cm}^3$ is the critical density,
and $h$ is the dimensionless Hubble parameter.
Before we scan the parameter space, we consider a representative point
in the four dimensional parameter space with $Q=1/3$:
\be
y_3&=& 0.00424, y_1=y_2=0.00296,~\lambda_S=0.13,
 \lambda_{HS}=0.06, \lambda_H=0.135~,\label{parameter}
 \ee
which gives
\be
& & \Omega h^2 = (\Omega_1+4 \Omega_2+3 \Omega_3)h^2=
0.119~,m_S = 324.1~\mbox{GeV},~\nn\\
& &m_1 =m_{\tilde{\eta}} =202.0~\mbox{GeV},~
m_2 = m_{\tilde{K}}=196.3~\mbox{GeV},~
m_3 = m_{\tilde{\pi}}=178.1~\mbox{GeV},~\nn\\
& &<\!v\sigma (11,22,33;X X) \!> 
=(9.29,~9.38,~1.26)\times 10^{-11} ~\mbox{GeV}^{-2}~,\nn\\
& &<\!v\sigma (11;22) \!> =4<\!v\sigma (22;11) \!>=3.90\times 10^{-5} ~\mbox{GeV}^{-2}~,\nn\\
& &<\!v\sigma (11;33) \!> =3 <\!v\sigma (33;11) \!>=4.30\times 10^{-5} ~\mbox{GeV}^{-2}~,\nn\\
& &<\!v\sigma (22;33) \!>= 
(3/4)<\!v\sigma (33;22) \!>=4.06\times 10^{-5}~ \mbox{GeV}^{-2}~,\nn\\
& &<\!\Gamma (1;\gamma\gamma)\!> 
= 6.45\times 10^{-13} ~\mbox{GeV}^{-1}~,\nn\\
& &<\!v\sigma (11;\gamma\gamma) \!>= 
6.59\times 10^{-14}~ \mbox{GeV}^{-2}
= 7.73\times 10^{-31}~ \mbox{cm}^{3}\mbox{s}^{-1}~.\nn
\ee
 Fig.~\ref{omegas} (left) shows  $\Omega h^2~(\mbox{red}),~
 \Omega_1 h^2=
 \Omega_\eta h^2~(\mbox{black}),~
4\Omega_2 h^2=4\Omega_K h^2 ~(\mbox{green})$ and 
$3\Omega_3 h^2=3\Omega_\pi h^2~(\mbox{blue})$ for the parameter values (\ref{parameter})
as a function of the inverse temperature $x=\mu/T$.
In Fig.~\ref{omegas} (right)  we show 
the total relic DM abundance $\Omega h^2$ as a function
of $y_1(=y_2)$,
where the other parameters are fixed as  (\ref{parameter}).
Since  a realistic  value of $\Omega h^2$ for the $SU(3)_V$  case (\ref{su3})
can be obtained only near the resonance, i.e. $m_S/m_{\rm DM}\simeq 2$,
the parameter space for the 
$SU(2)_V$ case (\ref{su2u1})  is 
considerably larger than that for the $SU(3)_V$ case  (\ref{su3}).
Note, however,   that  the realistic parameter space
for the $SU(2)_V$ case is not continuously connected 
to that for the $SU(3)_V$ case, 
 as we can see from Fig.~\ref{omegas} (right)
($SU(3)_V$ means the point at $y_1=y_3$).

\subsection{Indirect and Direct Detection  of Dark Matter}

\subsubsection{Monochromatic \mbox{\texorpdfstring{$\gamma$}{gamma}}-ray line from DM annihilation}
As we can see from Fig.~\ref{pp-to-gg},
two DM particles can  annihilate into two $\gamma$s.
Therefore, the charge $Q$ of the hidden fermions
can be constrained from the 
$\gamma$-ray observations
\cite{Ackermann:2012qk,Gustafsson:2013fca,Abramowski:2013ax}.
Since in the $SU(2)_V$ case  
the relic abundance of the $\tilde{\pi}$ dark matter is dominant,
we consider here only its annihilation into two $\gamma$s.
We will take into account only the s-wave contribution to
the annihilation cross section, and correspondingly we assume
that $p=p'=(m_{\tilde{\pi}},\bf 0)$ and that 
the photon momenta take the form
$k=(m_{\tilde{\pi}},\bf k)$ and $k'=(m_{\tilde{\pi}},-\bf k)$ with
their polarization tensors
 $\epsilon(k)=(0,\mbox{\boldmath  $\epsilon$}(k))$ and 
$\epsilon(k')=(0, \mbox{\boldmath  $\epsilon$}(k'))$ satisfying
 \be
0&=& \epsilon(k)\cdot k=
\epsilon(k)\cdot k'=
\epsilon(k)\cdot p=
\epsilon(k)\cdot p'~,\nn\\
0&=&
\epsilon(k')\cdot k=
\epsilon(k')\cdot k'=
\epsilon(k')\cdot p=\epsilon(k')\cdot p'~,
\ee
respectively.

To compute the annihilation rate we use the effective interaction 
(\ref{Lp2G2}).
We find that the annihilation amplitude can be written as
\be
\Gamma_{\mu\nu}(a~b)&\simeq & 
G_{\pi^2\gamma^2}
 \left( k\cdot k'  g_{\mu\nu}-k_\mu k'_\nu \right)
\times \left\{\begin{array}{cc} & a~b\\
1 & \gamma\gamma\\
-t_W&\gamma ~Z  \\
t_W^2&Z ~Z
 \end{array}~,
\right.
\label{Gmunu}
\ee
where $G_{\pi^2\gamma^2}$ is given in (\ref{p2g2}).
\begin{figure}
  \includegraphics[width=8cm]{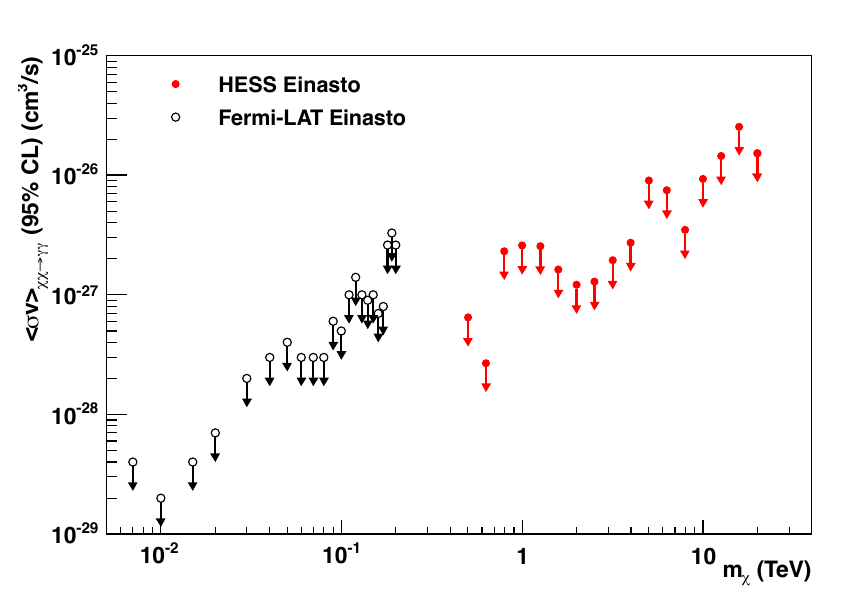}
 \includegraphics[width=7.5cm]{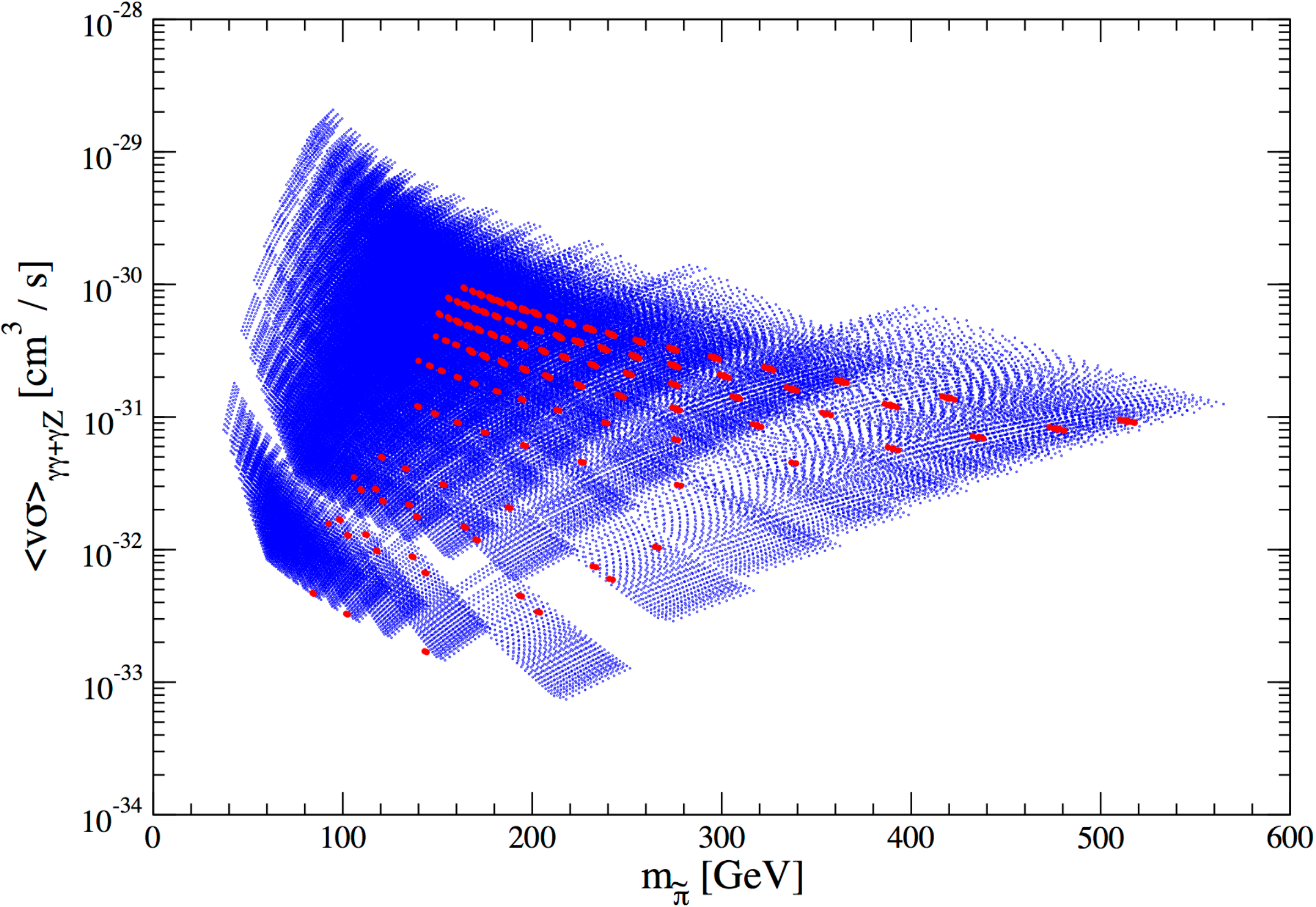}
\caption{\label{dmdmgg}\footnotesize
Left: The Fermi Lat \cite{Gustafsson:2013fca} (black)
and HESS\cite{Ackermann:2012qk} (red) 
upper bounds  on the velocity-averaged  DM annihilation
cross section for monochromatic $\gamma$-ray lines, where this graph is taken from  \cite{Ackermann:2012qk}.
Right: The velocity-averaged  DM annihilation cross section
$\langle v\sigma\rangle_{\gamma\gamma+\gamma Z}$
as a function of $m_{\tilde{\pi}}$
 with $Q=1/3$. Since $\langle v\sigma\rangle_{\gamma\gamma+\gamma Z}$ is proportional to
$Q^4$,  our calculations can be simply 
extended to the case of an arbitrary $Q$.
The red points are those for the $SU(3)_V$  case (\ref{su3}).}
\end{figure}
Then  (the s-wave part of) the corresponding   velocity-averaged  
annihilation cross sections are 
\be
\langle v\sigma(\tilde{\pi}\tilde{\pi}\to a~b)\rangle &= & 
\frac{G_{\pi^2\gamma^2}m_{\tilde{\pi}}^2}{4\pi }
\times \left\{\begin{array}{cc} & a~b\\
(1/2)& \gamma~\gamma \\
t_W^2(1-m_Z^2/4 m_{\tilde{\pi}}^2) &\gamma ~Z \\
(3/4)t_W^4(1-m_Z^2/m_{\tilde{\pi}}^2)^{1/2} & Z~Z
 \end{array}~.
\right.
\label{eq:gammaann}
\ee
The  energy $E_\gamma$ of $\gamma$-ray line produced 
by the annihilation into $\gamma Z$
is $m_{\tilde{\pi}}(1-m_Z^2/4 m_{\tilde{\pi}}^2)$.
In practice, however, due to finite detector energy resolution
this line cannot be distinguished 
from the $E_\gamma=m_{\tilde{\pi}}$ line.
Therefore, we simply add both cross sections.
So we compute
$\langle v\sigma\rangle_{\gamma\gamma+\gamma Z}
=\langle v\sigma(\tilde{\pi}\tilde{\pi}\to \gamma\gamma)\rangle+
\langle v\sigma(\tilde{\pi}\tilde{\pi}\to \gamma Z )\rangle$
with $Q=1/3$
as  a function  of $m_{\tilde{\pi}}$ for different values of $\lambda_H,~
\lambda_S$ and $\lambda_{HS}$, which is shown in 
Fig.~\ref{dmdmgg} (right), where 
$\Omega h^2$ 
 is required to be consistent with 
the PLANCK  experiment at $4\sigma$ level \cite{Planck:2015xua}.
As we see from Fig.~\ref{dmdmgg} (right)
the velocity- averaged annihilation cross section is mostly less than
$10^{-29}~\mbox{cm}^3/\mbox{s}$ in the parameter space we 
are considering, and consequently 
the Fermi LAT and HESS  constraints
given in Fig.~\ref{dmdmgg} (left) are well satisfied.
The red points are those for the $SU(3)_V$ (\ref{su3}) case.

The differential $\gamma$-ray flux is given by
\be
\frac{d\Phi}{d E_\gamma} &\propto &
 \langle v\sigma\rangle_{\gamma\gamma}
\frac{d N^{\gamma \gamma}}{d E_\gamma} +
\langle v\sigma\rangle_{\gamma Z}
\frac{d N^{\gamma z}}{d E_{\gamma Z}} 
\simeq \langle v\sigma\rangle_{\gamma\gamma+\gamma Z}
~\delta(E_\gamma-m_{\rm DM})~.
\ee
Prospects observing such line spectrum is discussed in detail in
\cite{Bringmann:2007nk,Bertone:2009cb,Laha:2012fg}. Obviously,   with an increasing energy resolution 
the  chance for the observation increases.
Observations of  monochromatic $\gamma$-ray lines of energies
 of $O(100)$ GeV 
 not only fix the charge of the hidden sector fermions, but also yields
 a first experimental hint on the hidden sector.

\subsubsection{Direct Detection  of Dark Matter}

\begin{figure}
 \includegraphics[width=4.5cm]{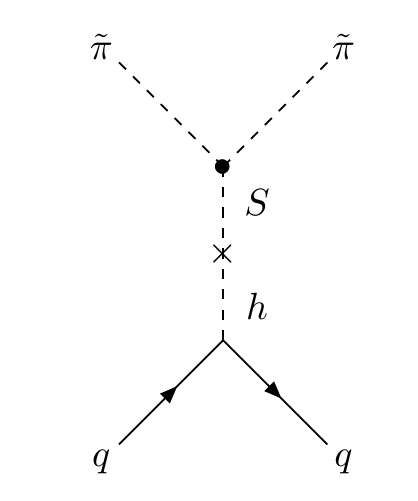}
  \includegraphics[width=4cm]{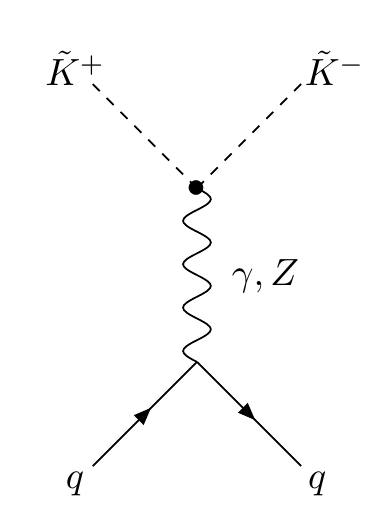}
 \includegraphics[width=7.5cm]{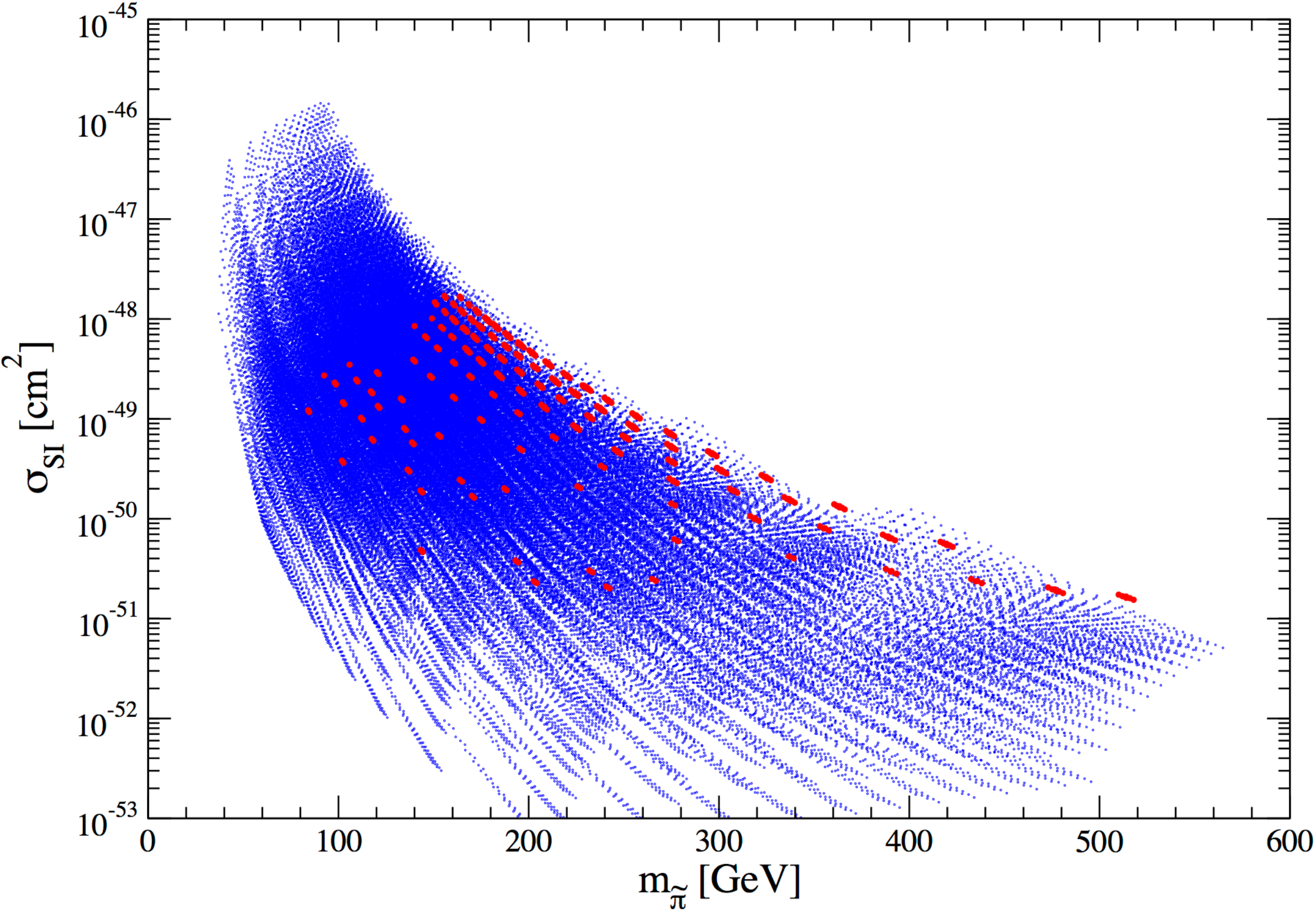}
\caption{\label{sigma}\footnotesize
Left: Subdiagrams contributing to 
the spin-independent elastic cross section $\sigma_{SI}$ 
 off   the nucleon. Since the relic abundance of $\tilde{K}$
 is negligibly small, the right diagram does not contribute.
Right: The spin-independent elastic cross section $\sigma_{SI}$ 
of $\tilde{\pi}$  as a function of $m_{\tilde{\pi}}$.
The red points are those for the $SU(3)_V$  case (\ref{su3}).
The result should be compared with the 
 XENON1T sensitivity of $O(10^{-47})~\mbox{cm}^2$ \cite{Aprile:2012zx}.}
\end{figure}

As we can see from Fig.~\ref{omegas} (left), 
the relic abundance of the $\tilde{K}$ dark matter is about 
three orders of magnitude smaller than that of the $\tilde{\pi}$ dark matter.
Therefore, we consider only the spin-independent 
elastic cross section $\sigma_{SI}$ of $\tilde{\pi}$ off the nucleon. The subprocess is the left diagram  
in Fig.~\ref{sigma} (left),
where $\bullet$ is the localized one-loop contribution (\ref{Gp2S}),
and we ignore the right diagram.
The result of \cite{Barbieri:2006dq} can be used to find
\be
\sigma_{SI}
&=&\frac{{Z_{\tilde{\pi}}^2}}{\pi} G_{\pi^2 S}^2
\left[ \frac{\hat{f}
m_N }{2 v_h m_{\tilde{\pi}}}\frac{\sin 2\theta}{2}
\left(\frac{1}{m_h^2}-
\frac{1}{m_S^2}\right) \right]^2
\left(\frac{m_N m_{\tilde{\pi}}}{m_N+m_{\tilde{\pi}}}
\right)^2~,
\label{sigmaSI}
\ee
where $G_{\pi^2 S}$ is given in
(\ref{Gp2S}), $m_N$ is the nucleon mass, and
$\hat{f}\sim 0.3$ stems from the nucleonic matrix element 
\cite{Ellis:2000ds}. We assume $|\cos\theta| \gtrsim  0.9$
to satisfy the LHC constraint, where  $\theta$ is the $h-S$ mixing
angle.
In Fig.~\ref{sigma} (right) we show in the  
$m_{\tilde{\pi}}\mbox{-}\sigma_{SI}$ plane
the area in which $\Omega h^2=0.12\pm 0.01
~(4\sigma)$    \cite{Planck:2015xua} is satisfied.
The predicted values of  $\sigma_{SI}$
 for $m_{\tilde{\pi}}  \gsim 150$ GeV is too small
 even for the future direct DM detection experiment 
 such as XENON1T, whose sensitivity is of $O(10^{-47})$
 $\mbox{cm}^2$ \cite{Aprile:2012zx}.
The smallness of $\sigma_{SI}$ results from the smallness of
the coupling $G_{\pi^2S}$, whose smallness comes from
small Yukawa coupling $y_1$ and the accidental cancellation
between the left and right diagrams in Fig.~\ref{pp-s}.
The red points are those for the $SU(3)_V$ (\ref{su3}) case.
We recall   that  the realistic parameter space
for the $SU(2)_V$ case is not continuously connected 
to that for the $SU(3)_V$ case,  as one could see from 
Fig.~\ref{omegas} (right),
in which   $y_1=y_3$ has to be satisfied for  the $SU(3)_V$ case.

If the relic abundance of the $\tilde{K}$ dark matter  were of $O(0.1)$,
the non-zero 
$\tilde{K}^0\mbox{-}\tilde{\bar{K}}^0\mbox{-}\gamma/Z$
and  $\tilde{K}^+\mbox{-}\tilde{K}^- \mbox{-}\gamma /Z$ couplings
shown in  Fig.~\ref{sigma} would lead to a serious problem. Fortunately,
the effective coupling is very small as we have already noticed:
$m_{\tilde{K}}^2 G_{K^+K^-\gamma}
\sim 10^{-6}$,
where this coupling for $\tilde{\pi}$ vanishes in the $SU(2)_V$
case.

Note that because 
an accidental $U(1)_V$  (the hidden baryon number),
not only the DM candidates, but also the lightest hidden baryons 
are stable. 
The hidden mesons in our model are neutral, while the charge of the hidden baryons 
$\tilde{b}$ formed by three hidden fermions is $Q_{\tilde{b}}=3\times Q$.
Let us roughly estimate the amount of relic stable hidden baryons and anti-baryons in the Universe, where we assume that the hidden proton and neutron are the lightest baryons
in the $SU(2)_V$ case (\ref{su2u1}).
As the hidden sector is  described by a scaled-up QCD, 
the hidden meson-baryon coupling $G_{\phi B \bar{B}}$ is approximately the 
same as in QCD, i.e.  $G_{\phi B \bar{B}}\sim 13$, which is independent of
$Q$. Using this fact, 
we can estimate $Y_{\tilde{b}}=n_{\tilde{b}}/s$ and obtain
$Y_{\tilde{b}} ~\simeq (0.4,~  6, ~9) \times 10^{-16}$ for
$m_{\tilde{b}}=1,~5$ and $8$ TeV, respectively.
There are severe constraints on $Y_{\tilde{b}}$.
The most severe constraints  exist for $Q_{\tilde{b}}=1$, which  come from 
the search of heavy isotopes in  sea water \cite{Agashe:2014kda} and also
from its influence on the large scale structure formation of the Universe 
\cite{Kohri:2009mi}. 
We therefore conclude that $Q_{\tilde{b}}=1$, i.e. $Q=1/3$ is ruled out.
Another severe cosmological constraint is due to
catalyzed BBN \cite{Pospelov:2006sc}, which gives
$Y_{\tilde{b}} \lsim 2\times 10^{-15}$ \cite{Pospelov:2008ta}
(see also   \cite{Hamaguchi:2007mp} ).
The CMB constraint based on the Planck data is 
$\Omega_{\tilde{b}} h^2 \lsim 0.001$ \cite{Dolgov:2013una},
which can be  satisfied in our model if $m_{\tilde{b}} \lsim 6$  TeV.
(See also \cite{Langacker:2011db} in which the constraints in the
$Q_{\tilde{b}}$-DM mass  plane are given,
where these constraints are satisfied in a wide area of the  parameter space of
the present model.). 

In most of our analyses on DM here we have used $Q=1/3$.
The relic abundances of DMs depend on $Q$, because the decay rate of the neutral
would-be DM $\eta$ depends on $Q$.
The change  of $Q$ can be compensated by
varying the ratio of $y_3$ to $y_1=y_2$,
as far as the difference of two hypercharges are not very much different.
As for the indirect detection of DM,
the annihilation cross section into 
two $\gamma$s  (\ref{eq:gammaann}) being proportional to $Q^4$ should be multiplied with $(3Q)^4$ for $Q$ different from $1/3$.
The spin-independent elastic cross section $\sigma_{SI}$ 
(\ref{sigmaSI}) is independent on $Q$.
This means that our basic results obtained in this
paper can be simply extended to the case with $Q$ different from $1/3$.

\section{Conclusion}

We have considered a  QCD-like hidden sector model
\cite{Hur:2007uz,Hur:2011sv,Heikinheimo:2013fta,Holthausen:2013ota}, in which dynamical chiral symmetry breaking
generates  a mass scale.
This generated scale is transmitted to the SM sector via a real SM singlet 
scalar $S$ to trigger spontaneous breaking of  EW  gauge symmetry
\cite{Hur:2007uz,Hur:2011sv}. Because the SM is extended  in a classically scale invariant way,
 "Mass without mass" \cite{wheeler,wilczek} is realized in this model.
Since chiral symmetry is dynamically broken,
there exist  NG bosons, which are massive
because the coupling of 
 $S$ with the hidden sector fermions breaks explicitly 
  the $SU(n_f)_L\times SU(n_f)_R$ chiral symmetry
  down to one of its diagonal subgroups.
 The mass scale of these NG bosons  is calculable 
 once the strength of this coupling and
 the scale of the QCD-like hidden sector are given.
The smallest  subgroup is 
the Cartan subalgebra $U(1)^{n_f-1}$. Because
of this (accidentally) unbroken subgroup, the NG bosons charged under  $U(1)^{n_f-1}$ are
stable: There exist at least $ n_f^2-n_f $ DM candidates.
We have restricted ourselves to  $n_c=n_f=3$, because in this case
we can relate using hadrons the independent parameters 
of the NJL model,
which we have used as a low-energy effective theory
for the hidden sector.
There are three possibilities:
(i) $U(1)_{\tilde{B}'}\times U(1)_{\tilde{B}}$, 
(ii) $SU(2)_V\times U(1)_{\tilde{B}}$, and
(iii) $SU(3)_V$, where the possibility (iii) has been studied in \cite{Holthausen:2013ota,Kubo:2014ida}.
It turns out that the first case (i) is unrealistic, 
unless  this case is very close to (ii) or (iii),
 or/and the hypercharge $Q$ of the hidden fermions
is tiny. This is  because the lightest NG boson  is neutral under
 $U(1)_{\tilde{B}'}\times U(1)_{\tilde{B}}$ so that it can decay 
 into two $\gamma$s  and the stable DM candidates annihilate  into them immediately. Therefore, we have mainly studied the case (ii)
 with $y_1=y_2 < y_3$.
 In this case the unstable NG boson is
$\tilde{\eta}$ (the heaviest among the pseudo NG bosons)
 and can decay into two $\gamma$s.
The annihilation cross section into the SM particles via the singlet $S$
is very much suppressed, except in the  resonance region
in the s-channel annihilation diagram of DM.
However, we have found another mechanism for the stable DMs
to annihilate:
 If $y_3$ does not differ very much from $y_1=y_2$,  the differences
among $m_{\tilde{\pi}}, m_{\tilde{K}}$ and $m_{\tilde{\eta}}$ are small.
At finite temperature the inverse DM conversions
(which are kinematically forbidden at zero temperature)
can become operative, because the DM conversions cross sections
are large $\sim 10^{-5}~\mbox{GeV}^{-2}$.
Consequently, the realistic  parameter space of the case (ii) is
significantly larger than that of  the case (iii), which has been
obtained in \cite{Holthausen:2013ota,Kubo:2014ida}.

With a non-zero $Q$ the hidden sector is doubly connected with the 
SM sector; we have a bright hidden sector at hand. 
The connection via photon and $Z$ opens possibilities to 
probe the hidden sector at collider experiments such as $e^+ e^-$ 
collision \cite{Fujii}. 
In particular, the would-be DM, $\tilde{\eta}$,  can decay into two $\gamma$s, which
would give a smoking-gun event.

\vspace{0.5cm}
\noindent{\bf Acknowledgements:} 
We thank S.~Matsumoto and H.~Takano  for useful discussions.
The work of M.~A. is supported in part by the Grant-in-Aid for Scientific 
Research (Grant Nos. 25400250 and 26105509).

\appendix

\section{The NJL Lagrangian in the self-consistent mean field 
(SCMF) approximation}
Here we consider the NJL Lagrangian ${\cal L}_{\rm NJL}$
 (\ref{eq:NJL10}) in the SCMF approximation \cite{Hatsuda:1994pi}.
In the SCMF approximation one splits up the NJL Lagrangian
 (\ref{eq:NJL10}) into the sum
\be
\mathcal{L}_{\rm NJL} &= &\mathcal{L}_{0}+\mathcal{L}_{I}~,
\ee
where  $\mathcal{L}_{I}$ is 
normal ordered 
(i.e. $\langle 0\vert \mathcal{L}_{I}\vert 0\rangle =0$), and $\mathcal{L}_{0}$ contains at most fermion bilinears
which are not normal ordered.
We find that  ${\cal L}_0$ can be written as
\begin{eqnarray}
{\cal L}_{0} &= &{\cal L}_{K}+{\cal L}_{D}+{\cal L}_{M}~,
\label{L0}
\end{eqnarray}
where 
\begin{eqnarray}
{\cal L}_{K}&=&{\rm Tr} {\bar{\psi}}(i {\gamma}^\mu \partial_{\mu}
+g' Q {\gamma}^\mu B_{\mu}){\psi}-\Bigl(\tilde{\sigma}_1+y_1 S-\frac{G_D}{8G^2}\tilde{\sigma}_2 \tilde{\sigma}_3 \Bigr)\bar{\psi_1}\psi_1\nn\\
& &-\Bigl(\tilde{\sigma}_2+y_2 S-\frac{G_D}{8G^2}\tilde{\sigma}_1 \tilde{\sigma}_3 \Bigr)\bar{\psi_2}\psi_2
 -\Bigl(\tilde{\sigma}_3+y_3 S-\frac{G_D}{8G^2}\tilde{\sigma}_1 \tilde{\sigma}_2 \Bigr)\bar{\psi_3}\psi_3\nn\\
& &-i\Bigl[\tilde{\pi}^0+\frac{1}{\sqrt3}\tilde{\eta}^8 +\sqrt{\frac{2}{3}}\tilde{\eta}^0 
-\frac{G_D}{8G^2}\Bigl (\tilde{\sigma}_3 \tilde{\pi}^0+\frac{1}{\sqrt3}( 2\tilde{\sigma}_2- \tilde{\sigma}_3)\tilde{\eta}^8 -\sqrt{ \frac{2}{3}} (\tilde{\sigma}_2+ \tilde{\sigma}_3)\tilde{\eta}^0\Bigr)\Bigr]\bar{\psi_1}{\gamma}_5 \psi_1\nn\\
 & &-i\Bigl[-\tilde{\pi}^0+\frac{1}{\sqrt3}\tilde{\eta}^8+\sqrt{\frac{2}{3}}\tilde{\eta}^0
-\frac{G_D}{8G^2} \Bigl(-\tilde{\sigma}_3 \tilde{\pi}^0+\frac{1}{\sqrt3}( 2\tilde{\sigma}_1- \tilde{\sigma}_3)\tilde{\eta}^8 -\sqrt{ \frac{2}{3}} (\tilde{\sigma}_1+ \tilde{\sigma}_3)\tilde{\eta}^0)\Bigr)\Bigr]\bar{\psi_2}{\gamma}_5 \psi_2
 \nn\\
& &-i\sqrt2\tilde{\pi}^+\Bigl(1-\frac{G_D}{8 G^2}\tilde{\sigma}_3\Bigr){\bar{\psi_1}}{\gamma}_5 \psi_2 
 -i\sqrt2 \tilde{\pi}^-\Bigl(1-\frac{G_D}{8 G^2}\tilde{\sigma}_3\Bigr){\bar{\psi_2}}{\gamma}_5 \psi_1
\nonumber\\
& &  -i\sqrt 2 \tilde{K}^+\Bigl(1-\frac{G_D}{8 G^2}\tilde{\sigma}_2\Bigr){\bar{\psi_1}}{\gamma}_5 \psi_3 
-i\sqrt 2 \tilde{K}^-\Bigl(1 -\frac{G_D}{8 G^2}\tilde{\sigma}_2\Bigr){\bar{\psi_3}}{\gamma}_5 \psi_1\nn\\
 & & -i \sqrt 2\tilde{K}^0 \Bigl(1-\frac{G_D}{8 G^2}\tilde{\sigma}_1\Bigr) {\bar{\psi_2}}{\gamma}_5 \psi_3 
 -i\sqrt 2 \bar \tilde{K}^0 \Bigl(1 -\frac{G_D}{8 G^2}\tilde{\sigma}_1\Bigr){\bar{\psi_3}}{\gamma}_5 \psi_2
\\
& &-i\Bigl[-\frac{2}{\sqrt3}\tilde{\eta}^8+\sqrt{\frac{2}{3}}\tilde{\eta}^0
-\frac{G_D}{8 G^2}
\Bigl((\tilde{\sigma}_1- \tilde{\sigma}_2) \tilde{\pi}^0-\frac{1}{\sqrt3}(\tilde{\sigma}_1+ \tilde{\sigma}_2)\tilde{\eta}^8 -\sqrt{\frac{2}{3}}(\tilde{\sigma}_1+ \tilde{\sigma}_2)\tilde{\eta}^0\Bigr)
\Bigr] {\bar{\psi_3}}{\gamma}_5 \psi_3~,\nn
 \label{LK}\\
{\cal L}_{D}&=& \frac{G_D}{8G^2}\Biggl[\Bigl(2\tilde{K}^0 \bar{\tilde{K}^0}- \frac{2}{\sqrt3}\tilde{\pi}^0 \tilde{\eta}^8 +\frac{2}{3}({\tilde{\eta}^8})^2-\frac{2}{3} ({\tilde{\eta}^0})^2\Bigr)
\bar{\psi_1}\psi_1
-\Bigl(2\sqrt{\frac{2}{3}}\tilde{\pi}^+ \tilde{\eta}^8 + 2\tilde{K}^+ \bar{\tilde{K}^0}\Bigr)\bar{\psi_1}\psi_2
\nonumber\\
& &-\Bigl(2\sqrt{\frac{2}{3}}\tilde{\pi}^- \tilde{\eta}^8 + 2\tilde{K}^-{\tilde{K}^0}\Bigr)\Bar{\psi_2}\psi_1
+\Bigl(2\tilde{K}^+ \tilde{K}^- + \frac{2}{\sqrt3}\tilde{\pi}^0 \tilde{\eta}^8 +\frac{2}{3}({\tilde{\eta}^8})^2-\frac{2}{3} ({\tilde{\eta}^0})^2\Bigr)
\bar{\psi_2}\psi_2
\nonumber\\
& &
-\Bigl(2\tilde{\pi}^+ \tilde{K}^0+\sqrt2\tilde{K}^+(\tilde{\pi}^0-\frac{1}{\sqrt3}\tilde{\eta}^8)\Bigr)\bar{\psi_1}\psi_3
-\Bigl(2\tilde{\pi}^- \bar{\tilde{K}^0}+\sqrt2 \tilde{K}^- (\tilde{\pi}^0-\frac{1}{\sqrt3}\tilde{\eta}^8)\Bigr)\bar{\psi_3}\psi_1\nonumber\\
& &-\Bigl(2\tilde{\pi}^- \tilde{K}^+ -\sqrt2\tilde{K}^0(\tilde{\pi}^0+\frac{1}{\sqrt3}\tilde{\eta}^8\bigr)\Bigr)\bar{\psi_2}\psi_3
-\Bigl(2\tilde{\pi}^+ \tilde{K}^- -\sqrt2 \bar \tilde{K}^0 (\tilde{\pi}^0+\frac{1}{\sqrt3}\tilde{\eta}^8) \Bigr)\bar{\psi_3}\psi_2
\nonumber\\
& &+\Bigl(2\tilde{\pi}^+ \tilde{\pi}^- +({\tilde{\pi}^0})^2-\frac{1}{3}(\tilde{\eta}^8)^2 -\frac{2}{3}(\tilde{\eta}^0)^2       \Bigr)\bar{\psi_3}\psi_3
\nonumber\\
& &+\sqrt{\frac{2}{3}}\tilde{\eta}^0 
\Bigl\{(\tilde{\pi}^0+\frac{1}{\sqrt3}\tilde{\eta}^8)\bar{\psi_1}\psi_1+\sqrt2\tilde{\pi}^+{\bar{\psi_1}}\psi_2+ \sqrt 2 \tilde{K}^+ {\bar{\psi_1}}\psi_3 
+\sqrt2\tilde{\pi}^- {\bar{\psi_2}}\psi_1\nonumber\\
& &+(-\tilde{\pi}^0+\frac{1}{\sqrt3}\tilde{\eta}^8)\bar{\psi_2}\psi_2+ \sqrt 2 \tilde{K}^0 {\bar{\psi_2}}\psi_3 
+\sqrt 2 \tilde{K}^- {\bar{\psi_3}}\psi_1+\sqrt 2\bar{\tilde{K}^0} {\bar{\psi_3}}\psi_2
-\frac{2}{\sqrt3}\tilde{\eta}^8{\bar{\psi_3}}\psi_3\Bigr\}
\Biggr]~,
\label{LD}\\
{\cal L}_{M} &=&
-\frac{1}{8G}\Biggl(\sum_{i=1}^3\tilde{\sigma}_i^2 +2
(\tilde{\eta}^0)^2+4\tilde{\pi}^+  \tilde{\pi}^- +4\tilde{K}^+ \tilde{K}^- +4 \tilde{K}^0 \bar \tilde{K}^0+2(\tilde{\pi}^0)^2+2(\tilde{\eta}^8)^2 \Biggr)\nonumber \\
& &+\frac{G_D}{16G^3}\Biggl[\tilde{\sigma}_1\tilde{\sigma}_2 \tilde{\sigma}_3+\tilde{\sigma}_1\Bigl (2\tilde{K}^0 \bar \tilde{K}^0
+\frac{2}{3}(\tilde{\eta}^8)^2-\frac{2}{3}(\tilde{\eta}^0)^2-\frac{2}{\sqrt3}\tilde{\pi}^0\tilde{\eta}^8+ \sqrt{\frac{2}{3}}\tilde{\eta}^0\tilde{\pi}^0+\frac{\sqrt2}{3}\tilde{\eta}^0\tilde{\eta}^8\Bigr)
\nonumber\\ 
& &
+\tilde{\sigma}_2\Bigl(2\tilde{K}^+ \tilde{K}^- +\frac{2}{3}(\tilde{\eta}^8)^2-\frac{2}{3}(\tilde{\eta}^0)^2+\frac{2}{\sqrt3}\tilde{\pi}^0\tilde{\eta}^8- \sqrt{\frac{2}{3}}\tilde{\eta}^0\tilde{\pi}^0+\frac{\sqrt2}{3}\tilde{\eta}^0\tilde{\eta}^8\Bigr)\nonumber\\ 
& &+
\tilde{\sigma}_3\Bigl(2\tilde{\pi}^+  \tilde{\pi}^- +(\tilde{\pi}^0)^2-\frac{1}{3}(\tilde{\eta}^8)^2-\frac{2}{3}(\tilde{\eta}^0)^2-\frac{2\sqrt2}{3}\tilde{\eta}^0\tilde{\eta}^8\Bigr)
\Biggr]~.
\label{LM}
\end{eqnarray}
Here $\tilde{\eta}^0$ stands for $\phi_0$, and the meson fields are defined in (\ref{mesons}).

\section{Determination of the NJL parameters 
\mbox{\texorpdfstring{$G,\,G_D$}{G,GD}} and 
\mbox{\texorpdfstring{$\Lambda$}{Lambda}} \label{sec:appendix}}
As in \cite{Holthausen:2013ota} we apply the NJL Lagrangian (\ref{L0}) 
with $g'=0$ to describe the real hadrons,
where we assume $SU(2)_V$ and replace $y_i S$ by the current 
quark masses, i.e. $y_1 S=y_2 S\to m_u~,~y_3 S\to m_s$.
Then we compute the real meson masses 
$m_\pi, m_K, m_\eta,m_{\eta'} $ and decay constants $f_\pi, f_K$.

We obtain the following inverse meson propagators:
\begin{eqnarray}
\Gamma_{\pi^{\pm}}({p^2})&=&\Gamma_{{\pi^{0}}}(p^2)\nonumber\\&=&-\frac{1}{2G}+\frac{G_D}{8G^3}\sigma_3+\Bigl(1-\frac{G_D}{8G^2}\sigma_3\Bigr)^2 2n_c I_{\phi^2}^A (p^2,m_1,m_1)+\frac{G_D}{G^2} n_c I_{\phi^2}^B(m_3)~,\\
\Gamma_{K^{\pm}}({p^2})&=&\Gamma_{\bar{K^0}K^{0}}(p^2)\nonumber\\&=&-\frac{1}{2G}+\frac{G_D}{8G^3}\sigma_1 
+\Bigl(1-\frac{G_D}{8G^2}\sigma_1\Bigr)^2 2n_c I_{\phi^2}^A (p^2,m_1,m_3)+\frac{G_D}{G^2} n_c I_{\phi^2}^B(m_1)~,\\
\Gamma_{\eta}^8(p^2)&=&-\frac{1}{2G}+\frac{G_D}{6G^3}
(\sigma_1-\frac{1}{4}\sigma_3)+\frac{2}{3}\Bigl(1-\frac{G_D}{8G^2}(2\sigma_1-\sigma_3)\Bigr)^2 n_c
 I_{\phi^2}^A (p^2,m_1,m_1)
\nonumber\\&+&\frac{4}{3}\Bigl(1-\frac{G_D}{8G^2}\sigma_1\Bigr)^2 n_c I_{\phi^2}^A (p^2,m_3,m_3)+\frac{4G_D}{3G^2} n_c I_{\phi^2}^B(m_1)-\frac{G_D}{3G^2} n_c I_{\phi^2}^B(m_3)~,\\
\Gamma_{\eta}^0(p^2)&=&-\frac{1}{2G}-\frac{G_D}{12G^3} \Big(2\sigma_1+\sigma_3\Bigr)+\Bigl(1+\frac{G_D}{8G^2}(\sigma_1+\sigma_3)\Bigr)^2 \frac{4}{3} 
n_c  I_{\phi^2}^A (p^2,m_1,m_1)
\nonumber\\&+&\Bigl(1+\frac{G_D}{4G^2}\sigma_1\Bigr)^2 \frac{2}{3} n_c  I_{\phi^2}^A (p^2,m_3,m_3)
-\frac{2G_D}{3G^2} n_c\Big(2 I_{\phi^2}^B(m_1)+I_{\phi^2}^B(m_3)\Bigr)~,\\
\Gamma_{{\eta}^8{\eta}^0} (p^2)&=&\frac{\sqrt{2} G_D}{24G^3}(\sigma_1-\sigma_3)
+\frac{2\sqrt{2}}{3}\biggl(1-\frac{G_D}{8G^2}(2\sigma_1-\sigma_3)\biggr)
\biggl(1+\frac{G_D}{8G^2}(\sigma_1+\sigma_3)\biggr)
n_c I_{\phi^2}^A (p^2,m_1,m_1)
\nonumber\\
&-&\frac{2\sqrt{2}}{3}\biggl(1-\frac{G_D}{8G^2}\sigma_1\biggr)\biggl(1+{\frac{G_D}{8G^2}}(2\sigma_1) \biggr)n_c I_{\phi^2}^A (p^2,m_3,m_3)
\nonumber\\
&+&\frac{\sqrt{2} G_D}{3G^2} n_c\bigl( I_{\phi^2}^B(m_1)- I_{\phi^2}^B(m_3)\bigr)~,
\end{eqnarray}
where  
the integrals $I_{\phi^2}^A (p^2,m_a,m_b)$ and $I_{\phi^2}^B(m)$ are defined in appendix (\ref{invP}), and
\begin{eqnarray}
m_1=m_u+\sigma_1-\frac{G_D}{8G^2}{\sigma_1}{\sigma_3}
~,~m_3=m_s+\sigma_3-\frac{G_D}{8G^2}{\sigma_1}^2~.
\end{eqnarray}
The  pion and kaon masses are the zeros of the inverse propagators, i.e.
\begin{eqnarray}
\Gamma_{\pi^{\pm}}(p^2 = {m_{\pi}}^2)=0
~,~\Gamma_{K^{\pm}}(p^2 = {m_K}^2)=0~,
\end{eqnarray}
while the $\eta$ and $\eta'$ meson masses are obtained 
from the zero eigenvalues of 
the real part of the $\eta^8-\eta^0$ mixing matrix. The wave function renormalization constants can be obtained from
\be
{Z_{\pi}}^{-1}&=&
\frac{d \Gamma_{\pi^{\pm}} (p^2)}{d p^2} {\biggl|}_{p^2={m_{\pi}}^2}~,~
{Z_{K}}^{-1}=
\frac{d \Gamma_{K^{\pm}} (p^2)}{d p^2} {\biggl|}_{p^2
={m_{K}}^2}~,
\label{wave}
\ee
and the pion and kaon decay constants are defined as 
\begin{eqnarray}
<0|{\rm Tr} \bar{\psi}{\gamma}_{\mu} 
{\gamma}_5\frac{1}{2}(\sigma_1+i \sigma_2)
{\psi}|{\pi^+}(p)> &=& i \sqrt{2} f_{\pi}p_{\mu}~,\\
<0|{\rm Tr} \bar{\psi}{\gamma}_{\mu} 
{\gamma}_5\frac{1}{2}(\sigma_4+i \sigma_5)
{\psi}|{K^+}(p)> &= & i \sqrt{2}  f_{K}p_{\mu}~.
\end{eqnarray}
We use
 $m_\pi, m_K, m_\eta, m_{\eta'}, f_\pi$ and $f_K$
to determine the QCD NJL parameters.
The best fit values of the parameters are given in Table \ref{tab:njlparameter}.
\begin{table}
\begin{tabular}{|c|c|c|c|c|c|} 
\hline
Parameter & $(2G^{\mathrm{QCD}})^{-1/2}$ & $(-G_D^{\mathrm{QCD}})^{-1/5}$ & $\Lambda^{\mathrm{QCD}}$  & $m_u$ & $m_s$\\ \hline
Value (MeV) & $361$ &$406$ &$930$ &
 $5.95$ & $163$ \\ \hline
\end{tabular}
\caption{Values of the QCD NJL parameters obtained by fitting the pion and kaon decay constants and the meson masses, where we have assumed
the $SU(2)_V$ flavor symmetry.}
\label{tab:njlparameter}
\end{table}
In Table \ref{tab:njlvalues} we compare the meson masses and decay constants calculated in the NJL theory with the experimental values.
\begin{table}
\begin{tabular}{|c|c|c|} 
\hline
  & Theory(MeV) & Experimental value(MeV) \\    \hline
$m_{\pi}$        &   $136$  &$140 (\pi^{\pm}) ~135(\pi^{0})$\\ 
$m_{K}$          &    $499$   & $494(K^{\pm}) ~498(K^{0},\bar{K^{0}})$\\ 
$m_{\eta} $     &    $460 $  & $548 $\\  
$m_{\eta'}$      &   $960$  & $958$ \\  
$f_{\pi}$        &      $93 $ & $92(\pi^{-})$ \\  
$f_{K}    $       &   $105 $ & $110(K^{-}) $ \\  \hline                       
\end{tabular}
\caption{Comparison of the NJL values with the corresponding
 experimental values. }
\label{tab:njlvalues}
\end{table}
As we see from Table \ref{tab:njlvalues}, the NJL $\eta$ mass
is  about $16$\% smaller than the experimental value.
This seems to be a general feature of the NJL theory
\cite{Hatsuda:1994pi} (see also \cite{Inagaki:2013hya}).

\section{One-loop Integrals}
\vspace{0.5cm}

\noindent
$\bullet$\underline{Vacuum  energy}\\
To compute the effective potential (\ref{eq:Vnjl}) we need the vacuum energy
\be
I_V(m)&=&\int\frac{d^4 k}{i(2\pi)^4}\ln \det (\slashed{k}-m) \nn \\ 
&=&\frac{1}{16\pi^2}\left(\Lambda^4 \ln\left(1+ \frac{m^2}{\Lambda^2} \right)
-m^4 \ln\left( 1+\frac{\Lambda^2}{m^2} \right)
+m^2 \Lambda^2\right)~.
\label{vac}
\ee

\noindent
$\bullet$\underline{Inverse propagator of dark matter}\\
There are two types of diagrams which contribute
to the inverse propagator of dark matter:
\be
I_{\phi^2}^A(p^2,m_a,m_b)&=&
\int \frac{d^4 l}{i(2\pi)^4}
\frac{\mbox{Tr}(\slashed{l}-\slashed{p}+m_a)
\gamma_5(\slashed{l}+m_b)\gamma_5}{((l-p)^2-m_a^2)
(l^2-m_b^2)},\nn \\
I_{\phi^2}^B(m)&=&\int \frac{d^4 k}{i(2\pi)^4}
\frac{m}{(k^2-m^2)}
= -\frac{1}{16\pi^2}m\left[\Lambda^2-m^2 \ln\left(1+\frac{\Lambda^2}{m^2} \right)\right]~.
\label{invP}
\ee
 These expressions are used to find DM masses and wave function renormalization
 constants given in (\ref{wave}), respectively.

\noindent
$\bullet$\underline{$\phi\mbox{-}\phi\mbox{-}\gamma$ amplitude}\\
\be
& &  I_{\phi^2\gamma}^{\mu}(p,p',m_a,m_b)\nn\\
& &=
(-1)\int \frac{d^4 l}{i(2\pi)^4}
\frac{\mbox{Tr}
(\slashed{l}+m_a)
\gamma_5(\slashed{l}-\slashed{p}'+m_b)\gamma^\mu
(\slashed{l}+\slashed{p}+m_b)\gamma_5}{((l+p)^2-m_b^2)
(l^2-m_a^2)((l-p')^2-m_b^2)}~
+(p\leftrightarrow p',m_a\leftrightarrow m_b),\nn\\
& &=-(p^\mu p'^\nu-p'^\mu p^\nu)
 (p+p')_\nu I_{\phi\gamma^2}(m_a,m_b)+\cdots
\ee
$\mbox{with}~p^2=p'^2$, where $\cdots$ stands for higher order terms in the 
expansion of the external momenta, and
\be
 I_{\phi^2\gamma}(m_a,m_b)
 &=&
 \frac{1}{8\pi^2}\frac{1}{(m_a^2-m_b^2)^2(m_a+m_b)^2}\nn\\
 & &\times\left(
 \frac{1}{2}(m_a-m_b) (m_a^3+5 m_a^2 m_b+5 m_a m_b^2+m_b^3)\right.\nn\\
 & & \left.
-  \frac{1}{3}(m_a^4+3 m_a^3 m_b+ m_a^2 m_b^2
+3 m_a m_b^3+m_b^4)\ln (m_a^2/m_b^2)^2\right)~.
\label{Ippg}
\ee
The effective $\phi\mbox{-}\phi\mbox{-}\gamma$ interaction
Lagrangian is given in (\ref{Lp2g}). 

\noindent
$\bullet$\underline{$\phi\mbox{-}\gamma\mbox{-}\gamma$ amplitude}\\
The $\phi(p)\mbox{-}\gamma(k)\mbox{-}\gamma(k')$ 
three-point function is needed to compute 
the decay $\tilde{\eta}$ into two $\gamma$s (Fig.~\ref{eta-gg}):
\be
& &  I_{\phi\gamma^2}^{\mu\nu}(k,k',m)\nn\\
& &=
(-1)\int \frac{d^4 l}{i(2\pi)^4}
\frac{\mbox{Tr}
(\slashed{l}-\slashed{k}'+m)
\gamma^\mu(\slashed{l}+m)\gamma^\nu
(\slashed{l}+\slashed{k}+m)\gamma_5}{((l+k)^2-m^2)
(l^2-m^2)((l-k')^2-m^2)}~
+(k\leftrightarrow k',\mu\leftrightarrow \nu)~\nn\\
& &=\frac{i}{4 \pi^2 m}\epsilon^{\mu\nu\alpha\beta}
 k_\alpha k'_\beta +\cdots
 \label{Ipgg}
\ee
The amplitude is thanks to $\gamma_5$ gauge invariant
even for a finite $\Lambda$, i.e.\\
$k_\nu I_{\phi\gamma^2}^{\mu\nu}(k,k',m)
=k'_\mu I_{\phi\gamma^2}^{\mu\nu}(k,k',m)=0$.
The amplitude without $\gamma_5$ 
correspond to the $S(p)\mbox{-}\gamma(k)\mbox{-}\gamma(k')$ 
three-point function, which we denote by 
$I_{0,S}^{\mu\nu}(k,k',m)$. This amplitude is not gauge invariant
so that we need  to apply  least subtraction method
 \cite{Kubo:2014ida}.
The subscript $0$ indicates that the amplitude is unsubtracted, and
we denote the subtracted gauge-invariant one by $I_{S}^{\mu\nu}(k,k',m)$.
In Appendix C we demonstrate how to use least subtraction method
for this case.

\noindent
$\bullet$\underline{$\phi\mbox{-}\phi\mbox{-}
\phi\mbox{-}\phi$ amplitude}\\
The $\phi(p)\mbox{-}\phi(p')\mbox{-}\phi'(k)\mbox{-}
\phi'(k')\mbox{-}$ 
four-point function is needed to compute 
the DM conversion cross section (diagrams of Fig.~\ref{conversion1}  ):
\be
& & I^A_{\phi^4}(p,p',k,k',m_a,m_b,m_c,m_d)  \nn\\
& &=(-1)\int \frac{d^4 l}{i(2\pi)^4}
\frac{\mbox{Tr}
(\slashed{l}+m_a)\gamma_5
(\slashed{l}-\slashed{p}'+m_b)\gamma_5
(\slashed{l}+\slashed{p}-\slashed{k}+m_c)\gamma_5
(\slashed{l}+\slashed{p}+m_d)\gamma_5}{(l^2-m_a^2)((l-p')^2-m_b^2)
((l+p-k)^2-m_c^2)((l+p)^2-m_d^2)}~\nn\\
& &+(p\leftrightarrow p',k\leftrightarrow k')~+
(p\leftrightarrow p')+(k\leftrightarrow k')~,\\
& & I^B_{\phi^4}(p,p',m_a,m_b)  \nn\\
& &=(-1)\int \frac{d^4 l}{i(2\pi)^4}
\frac{\mbox{Tr}
(\slashed{l}+m_a)
(\slashed{l}+\slashed{p}+\slashed{p}'+m_b)}{(l^2-m_a^2)((l+p+p')^2-m_b^2)}~. 
\ee
At the lowest order in the expansion in the external momenta
we obtain
\be
& & I^A_{\phi^4}(0,0,0,0,m_a,m_a,m_c,m_c)
=4I^{1A}_{\phi^4}(m_a,m_c)  \nn\\
& & =-\frac{1}{4\pi^2}\
\frac{m_a^2 \ln (\Lambda^2/m_a^2)
-m_c^2 \ln (\Lambda^2/m_c^2)}{(m_a^2-m_c^2)}+
\cdots~, \label{Ip41A}\\
& & I^A_{\phi^4}(0,0,0,0,m_a,m_a,m_a,m_d) 
=4 I^{2A}_{\phi^4}(m_a,m_d)  \nn\\
& &=-\frac{1}{4\pi^2}\left(\frac{m_a(m_a^2+m_a m_d-m_d^2) 
\ln (\Lambda^2/m_a^2)
-m_d^3 \ln (\Lambda^2/m_d^2)}{(m_a-m_d)(m_a+m_d)^2}\right.
\nn\\
& & \left.-\frac{m_a}{m_a+m_d}+\cdots\right),\label{Ip42A}\\
& & I^A_{\phi^4}(0,0,0,0,m_a,m_a,m_a,m_a)  
=4 I^{3A}_{\phi^4}(m_a) \nn\\
 & &=-\frac{1}{4\pi^2}\left(-1+\ln(\Lambda^2/m_a^2)+\cdots
 \right)~,\label{Ip43A}\\
& & I^B_{\phi^4}(0,0,m_a,m_b) 
=I^{1B}_{\phi^4}(m_a,m_b)  \nn\\
& &=\frac{-1}{4\pi^2}\left(-\Lambda^2+
\frac{m_a^3 \ln (\Lambda^2/m_a^2)
-m_b^3 \ln (\Lambda^2/m_b^2)}{m_a-m_b}+\cdots\right),\label{Ip41B}~\\
& & I^B_{\phi^4}(0,0,m_a,m_a) 
=I^{2B}_{\phi^4}(m_a)  \nn\\
& &=\frac{-1}{4\pi^2}\left(-\Lambda^2-
2m_a^2+3 m_a^2 \ln(\Lambda^2/m_a^2)+\cdots\right)~,\label{Ip42B}
\ee
where $\cdots$ stands for terms of $O(\Lambda^{-2})$
and higher. These expressions are used for the effective couplings
defined in (\ref{Ge2K2})-(\ref{GeK2p}).

\noindent
$\bullet$\underline{$\phi\mbox{-}\phi\mbox{-}S$ amplitude}\\
To obtain  the $\phi(p)\mbox{-}\phi(p')\mbox{-}S(k)$ 
three-point function (Fig.~\ref{pp-s}) we need
\be
& &  I_{\phi^2 S}^A(p,p',m_a,m_b)\nn\\
& &=
(-1)\int \frac{d^4 l}{i(2\pi)^4}
\frac{\mbox{Tr}
(\slashed{l}+p+m_b)
\gamma_5(\slashed{l}+m_a)\gamma_5
(\slashed{l}-\slashed{p}'+m_b)}{((l+p)^2-m_b^2)(l^2-m_a^2)
((l-p')^2-m_b^2)}~+(p\leftrightarrow p')~,\\
& &  I_{\phi^2 S}^B(p,p',m_a)\nn\\
& &=
(-1)\int \frac{d^4 l}{i(2\pi)^4}
\frac{\mbox{Tr}
(\slashed{l}+\slashed{p}+\slashed{p}'+m_a)
(\slashed{l}+m_a)}{((l+p+p')^2-m_a^2)(l^2-m_a^2)}~.
\ee

At the lowest order in the expansion in the external momenta
we obtain
\be
& & I_{\phi^2S}^{A}(0,0,m_a,m_b)=
2 I_{\phi^2S}^{1A}(m_a,m_b)   \nn\\
& & =\frac{1}{2\pi^2}\left(-\frac{m_b^2}{m_a+m_b}
-\frac{1}{2}(m_a-m_b) \ln (\Lambda^2/m_b^2)
+\frac{m_a^3}{2(m_a+m_b)^2} \ln (m_a^2/m_b^2)+
\cdots\right),\label{Ip2s1A}\\
& & I_{\phi^2S}^A(0,0,m_a,m_a)=
2I_{\phi^2S}^{2A}(m_a) \nn\\
& &
=\frac{m_a}{4\pi^2}\left(-1+ \ln(\Lambda^2/m_a^2)+\cdots\right)~,
\label{Ip2s2A}\\
& & I_{\phi^2S}^{B}(0,0,m_a)=
I_{\phi^2S}^{B}(m_a)   \nn\\
& &
=\frac{1}{4\pi^2}
\left(\Lambda^2+2 m_a^2-3 m_a^2\ln(\Lambda^2/m_a^2)+
\cdots\right)~.\label{Ip2sB}
\ee
These expressions are used for the effective couplings
defined in (\ref{Ge2S})-(\ref{Gp2S}).

\noindent
$\bullet$\underline{$\phi\mbox{-}\phi\mbox{-}S\mbox{-}S$ amplitude}\\
Similarly,
\be
& & I^A_{\phi^2S^2}(p,p',k,k',m_a,m_b,m_c,m_d)  \nn\\
& &=(-1)\int \frac{d^4 l}{i(2\pi)^4}
\frac{\mbox{Tr}
(\slashed{l}+m_a)\gamma_5
(\slashed{l}-\slashed{p}'+m_b)
(\slashed{l}+\slashed{p}-\slashed{k}+m_b)
(\slashed{l}+\slashed{p}+m_b)\gamma_5}{(l^2-m_a^2)((l-p')^2-m_b^2)
((l+p-k)^2-m_b^2)((l+p)^2-m_b^2)}~\nn\\
& &+(k\leftrightarrow k')~,\\
& & I^B_{\phi^2S^2}(p,p',k,k',m_a,m_b,m_c,m_d)  \nn\\
& &=(-1)\int \frac{d^4 l}{i(2\pi)^4}
\frac{\mbox{Tr}
(\slashed{l}+m_a)
(\slashed{l}+\slashed{k}'+m_a)\gamma_5
(\slashed{l}+\slashed{p}-\slashed{k}+m_b)
(\slashed{l}+\slashed{p}+m_b)\gamma_5}{(l^2-m_a^2)((l+k')^2-m_b^2)
((l+p-k)^2-m_b^2)((l+p)^2-m_b^2)}~\nn\\
& &+(k\leftrightarrow k')~,\\
& & I^C_{\phi^4}(p,p',m_a)  \nn\\
& &=
(-1)\int \frac{d^4 l}{i(2\pi)^4}
\frac{\mbox{Tr}
(\slashed{l}-\slashed{k}'+m)(\slashed{l}+m)
(\slashed{l}+\slashed{k}+m)}{((l+k)^2-m^2)
(l^2-m^2)((l-k')^2-m^2)}~
+(k\leftrightarrow k')~. 
\ee

At the lowest order in the expansion in the external momenta
we obtain
\be
& & I^A_{\phi^2S^2}(0,0,0,0,m_a,m_b)
=2 I^{1A}_{\phi^2S^2}(m_a,m_b)  \nn\\
& & =-\frac{1}{2\pi^2}\left[
\frac{1}{(m_a+m_b)^2}\left(
m_b(5 m_a+3 m_b)
+\frac{m_a^3}{(m_a+m_b)}\ln (m_a^2/m_b^2)\right)\right.\nn\\
& &\left.-\ln (\Lambda^2/m_a^2)+\cdots\right]~,
\label{Ip2s21A}\\
& & I^A_{\phi^2S^2}(0,0,0,0,m_a,m_a)=
2I^{2A}_{\phi^2S^2}(m_a)  \nn\\
& &=-\frac{1}{2\pi^2}\left(
2-\ln (\Lambda^2/m_a^2)
+\cdots\right)~,
\label{Ip2s22A}\\
& & I^B_{\phi^2S^2}(0,0,0,0,m_a,m_b)
=2 I^{1B}_{\phi^2S^2}(m_a,m_b)  \nn\\
& & =-\frac{1}{2\pi^2}
\frac{1}{(m_a+m_b)^2}\left(
\frac{m_a^2(m_a+3 m_b)\ln (\Lambda^2/m_a^2)+
m_b^2(m_b+3 m_a)\ln (\Lambda^2/m_b^2)
}{(m_a+m_b)}\right.\nn\\
& &\left.-2 ( m_a^2+ m_b^2)+\cdots\right)~,\label{Ip2s21B}\\
& & I^B_{\phi^2S^2}(0,0,0,0,m_a,m_a)=
2 I^{2B}_{\phi^2S^2}(m_a)  \nn\\
& &=-\frac{1}{2\pi^2}\left(
-1+\ln (\Lambda^2/m_a^2)
+\cdots\right)~,\label{Ip2s22B}\\
& & I^{C}_{\phi^2S^2}(0,m)
 = 2  I^{C}_{\phi^2S^2}(m)=\frac{m}{2\pi^2}\left(
5-3\ln (\Lambda^2/m^2)
+\cdots\right)~.\label{Ip2s2C}
\ee
These expressions are used for the effective couplings
defined in (\ref{Ge2S2})-(\ref{Gp2S2}).

\vspace{0.5cm}
\noindent
$\bullet$\underline{$\phi\mbox{-}\phi\mbox{-}\gamma\mbox{-}\gamma$ amplitude}\\
The next example is the 
$\phi(p)\mbox{-}\phi(p')\mbox{-}\gamma(k)\mbox{-}\gamma(k')$ four-point 
function.
The diagrams at the one-loop level are shown 
in Fig.~\ref{pp-to-gg}:
\be
& & I_{0,\phi^2}^{A,\mu\nu}(p,p',k,k',m_a,m_b,m_c)  \nn\\
& &=(-1)\int \frac{d^4 l}{i(2\pi)^4}
\frac{\mbox{Tr}
(\slashed{l}+m_a)\gamma_5
(\slashed{l}-\slashed{p}'+m_b)\gamma^\mu
(\slashed{l}+\slashed{p}-\slashed{k}+m_c)\gamma^\nu
(\slashed{l}+\slashed{p}+m_b)\gamma_5}{(l^2-m_a^2)((l-p')^2-m_b^2)
((l+p-k)^2-m_c^2)((l+p)^2-m_b^2)}~\nn\\
& &+(p\leftrightarrow p',k\leftrightarrow k',\mu\leftrightarrow \nu)+
(k\leftrightarrow k',\mu\leftrightarrow \nu)+
(p\leftrightarrow p')~,\\
& & I_{0,\phi^2}^{B,\mu\nu}(p,p',k,k',m_a,m_b)  \nn\\
& &=(-1)\int \frac{d^4 l}{i(2\pi)^4}
\frac{\mbox{Tr}
(\slashed{l}-\slashed{k}+m_a)\gamma^\nu
(\slashed{l}+m_a)\gamma_5
(\slashed{l}-\slashed{p}'+m_b)\gamma^\mu
(\slashed{l}+\slashed{p}-\slashed{k}+m_b)\gamma_5}
{((l-k)^2-m_a^2)(l^2-m_a^2)
((l-p')^2-m_b^2)((l+p-k)^2-m_b^2)}~\nn\\
& &+
(p\leftrightarrow p')~,\\
& &  I_{0,\phi^2}^{C,\mu\nu}(k,k',m)\nn\\
& &=
(-1)\int \frac{d^4 l}{i(2\pi)^4}
\frac{\mbox{Tr}
(\slashed{l}-\slashed{k}'+m)
\gamma^\mu(\slashed{l}+m)\gamma^\nu
(\slashed{l}+\slashed{k}+m)}{((l+k)^2-m^2)
(l^2-m^2)((l-k')^2-m^2)}~
+(k\leftrightarrow k',\mu\leftrightarrow \nu)~.
\ee
The subscript $0$ indicates that the amplitudes are
unsubtracted, and therefore they are not gauge invariant.
We apply  least subtraction method to obtain
gauge invariant amplitudes $I_{\phi^2}^{A,\mu\nu},
I_{\phi^2}^{B,\mu\nu} $ and $ I_{\phi^2}^{C,\mu\nu}$,
respectively. 
Since the realistic parameter space is  close to that of
the $SU(2)_V$ case (\ref{su2u1}), we consider them only 
in this case.
At the lowest order in the expansion in the external momenta
we obtain
\be
& &I_{\phi^2}^{A,\mu\nu}(k,k',m)+
I_{\phi^2}^{B,\mu\nu}(k,k',m)
=
\frac{1}{6 \pi^2 m^2}  \left( k\cdot k'  g^{\mu\nu}-k^\mu k'^\nu
 \right)+\cdots
  \label{Ip2g2A}\\ 
 & &I_{\phi^2}^{C,\mu\nu}(k,k',m) =
-\frac{1}{6 \pi^2 m} \left( k\cdot k'  g^{\mu\nu}-k^\mu k'^\nu
 \right)+\cdots
 \label{Ip2g2C}
 \ee
in the large $\Lambda$ limit.
The result is used for the effective Lagrangian (\ref{Lp2G2}) 
and (\ref{Gmunu}).

\section{Least Subtraction Procedure \label{sec:app}}

The cutoff $\Lambda$ breaks gauge invariance explicitly and to restore gauge invariance we have to subtract non-gauge invariant terms
from the original amplitude.
In  renormalizable theories there is no problem to define
a finite renormalized gauge invariant amplitude.
In  the limit of $\Lambda\to \infty$
the gauge non-invariant terms are a finite number of local 
terms,
 which can be cancelled by the corresponding
local counter terms so that the subtracted amplitude
is, up to its normalization,  independent of the regularization scheme.
To achieve such a uniqueness
in  cutoff theories, one needs 
an additional prescription.

In \cite{Kubo:2014ida}
 a novel method called
``\emph{least subtraction procedure}''  has been proposed. 
The basic idea is
to keep the subtraction terms to the minimum necessary.
Consider an unsubtracted amplitude 
\be
{\cal A}_{0,\mu_1\dots\mu_{n_g}}(\Lambda; 
p_1\dots p_{n_s},k_1\dots k_{n_g}), 
\label{Amunu}
\ee
 with $n_g$ photons and $n_s$ scalars (scalars and axial scalars). 
 Expand the amplitude in the external momenta
$k$'s and $p$'s:
\be
{\cal A}_{0,\mu_1\dots\mu_{n_g}}&=&
\sum_{m=0}{\cal A}^{(m)}_{0,\mu_1
\dots\mu_{n_g}}~, 
\ee
where ${\cal A}^{(m)}_{0,\mu_1\dots\mu_{n_g}}$
consists of  $m$-th order  monomials  of 
 the external momenta.
In general, 
${\cal A}^{(0)}_{0,\mu_1\dots\mu_{n_g}}
={\cal A}_{0,\mu_1\dots\mu_{n_g}}(\Lambda; 
0,\cdots,0)$
is non-vanishing and we can subtract it because it is not gauge invariant.
We keep the tensor structure
of ${\cal A}^{(0)}_{0,\mu_1\dots\mu_{n_g}}$
as the tensor structure of the counter terms for
${\cal A}^{(m)}_{0,\mu_1\dots\mu_{n_g}} ~(m>0)$
until a new tensor structure for the counter terms is required. 
We continue this until  no more new tensor structure  is needed.

To illustrate the subtraction method we consider 
the $S(p)\mbox{-}\gamma(k)\mbox{-}\gamma(k')$ 
three-point function,
which is given by
\be
{\cal A}_{0,\mu\nu}(k,k')& =&
\sum_{i=1}^3 y_i n_c e^2 Q^2\int \frac{d^4 l}{i(2\pi)^4}
\frac{\mbox{Tr}
(\slashed{l}-\slashed{k}'+M_i)
\gamma_\mu(\slashed{l}+M_i)\gamma_\nu
(\slashed{l}+\slashed{k}+M_i)}{((l+k)^2-M_i^2)
(l^2-M_i^2)((l-k')^2-M_i^2)}\nn\\~
& &+(k\leftrightarrow k',\mu\leftrightarrow \nu)~,
\ee
where we use the on shell conditions $k^2=k'^2=0$.
Without lost of generality the amplitude can be written as
\be
{\cal A}_{0,\mu\nu}(k,k') &=&
{\cal A}_{0,g} (k,k') g_{\mu\nu}+{\cal A}_{0,k} (k,k') k_\mu k'_\nu+
{\cal B}_{0,k}(k,k') k_\nu k'_\mu~.
\ee
The last term does not contribute to
the gauge invariance
$k^\nu {\cal A}_{\mu\nu}(k,k') 
=k'^\mu {\cal A}_{\mu\nu}(k,k') =0$, and so we ignore it.
The corresponding one-loop diagram
is the one in
Fig.~\ref{eta-gg} with $\tilde{\eta}$ replaced by $S$.
According to  least subtraction method, we expand the amplitude
in the external momenta $k$ and $k'$.
At the second order, for instance, we find
\be
{\cal A}_{0,g}^{(2)}(k,k') &=&
-\frac{n_c e^2 Q^2}{4\pi^2}(k\cdot k' )\sum_i\frac{y_i \Lambda^4}{3 M_i 
( \Lambda^2+M_i^2)^3}(2 \Lambda^2+M_i^2),~\\
{\cal A}_{0,k}^{(2)}(k,k') &= &
 \frac{n_c e^2 Q^2}{4\pi^2}\sum_i \frac{y_i \Lambda^4}{3 M_i ( \Lambda^2+ M_i^2)^3}(2 \Lambda^2+2M_i^2)~.
\ee
In the $\Lambda\to \infty$ limit the second order amplitude 
will be gauge invariant, but it is not at a finite $\Lambda$.
Moreover, there are infinitely many ways of subtraction
to make the  second order amplitude gauge invariant.
However, none of them is 
preferential. Least subtraction method
uses the lower order amplitude, i.e.  
\be
{\cal A}_{0,g}^{(0)}(k,k') &=&
-\sum_i\frac{\Lambda^4 M_i}{
( \Lambda^2+M_i^2)^2},~
{\cal A}_{0,k}^{(0)}(k,k') =0
\ee
in this case,  how to subtract
the second order amplitude. 
At the lowest order in the derivative expansion,
what is to be subtracted is unique; it is the  $g_{\mu\nu}$ term.
We keep this tensor structure
as the tensor structure of the counter terms for
higher order terms
until a new tensor structure for the counter terms is required. 
However, in the  case of ${\cal A}_{0,\mu\nu}(k,k')$
there will be  no new tensor structure appearing in higher orders.
This implies that  ${\cal A}_{0,k}(k,k')$ remains unsubtracted
(i.e. $ {\cal A}_{k}(k,k')={\cal A}_{0,k}(k,k') $) so that the subtracted gauge invariant amplitude is given by
\be
{\cal A}_{\mu\nu}(k,k')
&=&
-\sum_i \frac{y_i n_c e^2 Q^2}{4 \pi^2}\left(
g_{\mu\nu} k\cdot k'-k_\mu k'_\nu  \right)
\left( \frac{\Lambda^4}{M_i(\Lambda^2+M_i^2)^2} \right)\nn\\
& & \times\left[\frac{2}{3}+\frac{7k\cdot k' 
(\Lambda^2+3 M_i^2)}{90 M_i^2(\Lambda^2+M_i^2)}+
\frac{(k\cdot k' )^2
(\Lambda^4+4 \Lambda^2 M_i^2+6  M_i^4)}{63 M_i^4(\Lambda^2+M_i^2)^2}
+\cdots \right]\nn\\
&=&-\sum_i \frac{y_i n_c e^2 Q^2   M_i}{4\pi^2}
(g_{\mu\nu}k\cdot k'-k_\mu k'_\nu)\nn\\
& &
\times\int_0^1 dx\int_0^{1-x} dy~\frac{2\Lambda^4}{(\Lambda^2+D^2)^2}
\frac{(1-4 x y)}{D^2}~,
\label{Sgg}
\ee
where $D=M_i^2-2 x y k\cdot k' $.

\end{document}